\documentclass[%
 reprint,
 amsmath,amssymb,
 aps,apj
]{emulateapj}

\usepackage{amsmath}
\usepackage{amssymb}
\usepackage{amsfonts}
\usepackage{amssymb}
\usepackage{dcolumn}
\usepackage{bm}
\usepackage{bbm}
\usepackage{graphicx}
\usepackage{xcolor}
\usepackage{array}
\usepackage{subfigure}
\usepackage{wasysym}
\usepackage{graphicx}
\usepackage{hyperref}
\usepackage{cleveref}

\graphicspath{{Fig/}}

\def\Im{\mathrm{Im}\,}
\def\erfc{{\rm erfc}\,}

\definecolor{myblue}{RGB}{0, 100, 200}
\definecolor{myred}{RGB}{214, 39, 40}
\definecolor{mygreen}{RGB}{44, 160, 44}
\definecolor{mybrown}{RGB}{123, 64, 26}
\definecolor{mydarkblue}{RGB}{44, 77, 118}

\hypersetup{
	colorlinks=true,
	linkcolor=myblue,
	filecolor=blue,      
	urlcolor=blue,
	citecolor=myred,
}

\newcommand{\change}[1]{{\color{black}{#1}}}

\begin{document}

\title{Survival of the Fittest: \\Testing Superradiance Termination with Simulated Binary Black Hole Statistics}



\author{Hui-Yu Zhu$^{1,2}$}
\email{hzhuav@connect.ust.hk}
		
\author{Xi Tong$^{3}$}
\email{xt246@cam.ac.uk}
	
\author{Giorgio Manzoni$^{2}$}
\email{iasmanzoni@ust.hk}
	
\author{Yanjiao Ma$^{1,2}$}
\email{ymaby@connect.ust.hk}

\affiliation{${}^1$Department of Physics, The Hong Kong University of Science and Technology, \\ Clear Water Bay, Kowloon, Hong Kong, P.R.China}
	
\affiliation{${}^2$Jockey Club Institute for Advanced Study, The Hong Kong University of Science and Technology, Hong Kong S.A.R., China}
	
\affiliation{${}^3$Department of Applied Mathematics and Theoretical Physics, University of Cambridge,\\Wilberforce Road, Cambridge, CB3 0WA, UK}%

\begin{abstract}
The superradiance instability of rotating black holes leads to the formation of an ultralight boson cloud with distinctive observational signatures, making black holes an effective probe of ultralight \change{bosons}. However, around black holes in a binary system, the superradiance effect of such clouds can be terminated by tidal perturbations from the companion, leading to cloud depletion. In this study, we \change{focus on the superradiance of a scalar boson,} and perform the first analysis of the impact of this termination effect on superradiant black hole binaries which are \textit{realistically} modeled after their statistics in our Galaxy. Working with a dataset of approximately $10^7$ black hole binaries simulated using the Stellar EVolution for N-body (SEVN) population synthesis code, we identify the superradiant candidates and those that manage to survive the termination effect. We then calculate the cloud survival rate for various boson masses and black hole spin models. Our findings reveal that the $l=m=1$ cloud modes are generally stable against termination, whereas the $l=m=2$ modes can be significantly affected, with survival rates dropping below $10\%$ for boson masses below approximately $0.5\times 10^{-12}$ eV. In addition, our analysis indicates that clouds that overcome termination typically exhibit a higher superradiant growth rate and therefore a higher detectability.

\end{abstract}











\section{Introduction} 
\label{Sec.intro}

Light bosons are a natural extension of the Standard Model particle spectrum we know of today. For instance, the strong CP problem in quantum chromodynamics motivates the introduction of a light scalar particle known as the axion \citep{Peccei:1977hh,Weinberg:1977ma,Wilczek:1977pj,Kim:1979if}. In string theory, the dimensional reduction of gauge fields also leads to a plenitude of scalars with various light masses \citep{Svrcek:2006yi,Arvanitaki:2009fg,Mehta:2021pwf}. Interacting mostly through gravity while having a long de Broglie wavelength, these light axion-like scalars can also serve as wave dark matter and affect the large-scale structure of the universe \citep{Dine:1982ah,Preskill:1982cy,Abbott:1982af,Hui:2016ltb}.

\vskip 4pt

If such light scalar bosons exist near a rotating black hole (BH), the geometry of the BH can amplify the boson field by triggering an instability in the spectrum of the scalar boson bound states, a mechanism widely known as superradiance \citep{Detweiler:1980uk,Damour:1976kh,Brito:2015oca,Cannizzaro:2023jle}.\footnote{To be more precise, \change{the terminology superradiance may account for both superradiance scattering and superradiance instability. The former stands for the one-time amplification of scattering states, whereas the latter represents the long-term growth of bounds states. In this work, we focus on the latter case of superradiance instability.}} This leads to the formation of a boson cloud near the BH with a structure analogous to that of a Hydrogen atom, which is characterized by a set of ``quantum numbers'' $(nlm)$. Superradiance is most pronounced when the gravitational fine structure constant reaches \change{$\alpha\equiv M\mu\lesssim 1$}\footnote{\change{Throughout this paper, we set $G=\hbar=c=1$. }}, with $M$ being the mass of the BH and $\mu$ being that of the boson. For astrophysical BHs between $M\approx10^0$-$10^{10} M_\odot$, this maps to ultralight boson masses between $\mu\approx10^{-20}$-$10^{-10}\mathrm{\,eV}$. 

\vskip 4pt

The gravitational atom exhibits rich physics that has become widely explored in recent years. 
For instance, the annihilation process of boson pairs in the gravitational atom emits monochromatic gravitational waves (GW) \citep{Arvanitaki:2010sy,Yoshino:2013ofa,Chan:2022dkt,Banerjee:2024nga} that can fall into the sensitivity bands of various ground-based or space-based GW detectors \citep{Brito:2017zvb,Brito:2017wnc,Isi:2018pzk,Ghosh:2018gaw,Tsukada:2018mbp,LISAConsortiumWaveformWorkingGroup:2023arg}. In addition, due to the angular momentum extraction of the boson cloud, the BH spin continuously decreases to a \change{threshold} value beyond which the superradiance \change{for the low-$m$ modes} is no longer sustained\change{, while higher-$m$ modes are still superradiant}. This \change{threshold} spin is determined solely by the masses of the boson mass and the BH, making the observed spin-mass distribution of BHs a smoking gun of ultralight bosons \citep{Brito:2017zvb,Arvanitaki:2016qwi,Ng:2020ruv,Hui:2022sri,Ayzenberg:2023hfw}. Furthermore, given a specific underlying theory for the boson (e.g. the axion), the self-interaction among the different modes of the cloud can result in cloud dissipation \citep{Fukuda:2019ewf,Baryakhtar:2020gao,Omiya:2022mwv,Takahashi:2024fyq} and influence the GW signals \citep{Collaviti:2024mvh}. A statistical approach to the boson mass and cloud self-interaction is studied in \citet{Hoof:2024quk}. In a binary system, the dynamics become even more intriguing. When the orbital frequency matches the energy difference between two modes of the gravitational atom, a resonant transition can occur with backreaction on binary orbit \citep{Baumann:2019ztm,Baumann:2018vus,Baumann:2019eav}. Such resonances can be detected via multi-messenger probes \citep{Ding:2020bnl,Tong:2021whq}.

\vskip 4pt

Notice that all the phenomena above hinge on the existence of an adequately populated boson cloud. For an isolated BH, this is possible if the BH spin is large enough. For a BH in a binary system, the tidal perturbation from the companion mixes superradiant growing modes with non-superradiant decaying modes, thereby destabilizing the cloud \citep{Arvanitaki:2014wva,Du:2022trq} and \change{turning the effective superradiance rate to zero or below}. \change{This effect known as \textit{superradiance termination} occurs for generic binary orbits and tightly constrains the dynamics of boson clouds \citep{Tong:2022bbl}.} In Extreme-Mass-Ratio-Inspiraling (EMRI) systems, the backreaction of cloud termination can significantly alter the motion of the companion, \change{reducing the eccentricity and the misalignment angle between the BH spin and the orbital angular momentum} \citep{Fan:2023jjj}. Although the majority of BHs are expected to be isolated due to the kick they receive during the supernova explosion, the BHs observable through the emission of GWs are expected to be found in binary systems \citep{Lam:2022,Belczynski:2004,Fender:2013,Wiktorowicz:2019}. Moreover, fast-spinning BHs tend to acquire their high spin through the accretion of material from a close companion \citep{Gou:2011,McClintock:2014}, whereas isolated BHs are believed to be born with negligible spin \citep{Fuller:2019sxi}, making superradiance unlikely. Therefore, a systematic investigation of superradiance termination for realistic BH binaries is crucial to understanding superradiant boson clouds that could exist in our Galaxy.

\vskip 4pt

\begin{figure}[ht]
	\centering
	\includegraphics[width=8cm]{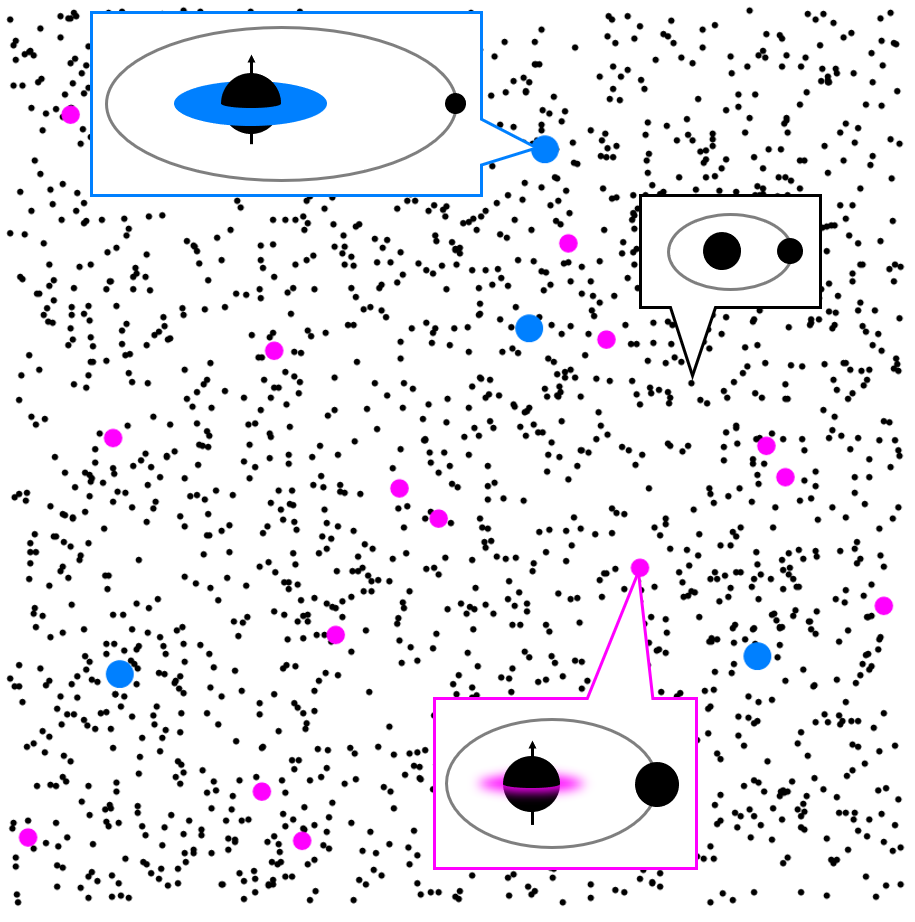}
	\caption{A schematic representation of the distribution of superradiant BH binaries in the galaxy. The coloured dots represent BH binaries that possess the potential of producing a superradiant cloud (participants) whereas the black dots represent those that do not. Among the coloured participants, those with a superradiant cloud that survives the tidal destabilization from the binary companion are marked in \textcolor{myblue}{blue} (survivors), and those whose superradiant clouds are terminated are marked in \textcolor{magenta}{magenta}. In this figure, 1\% of the BH binaries are capable of producing $\psi_{322}$ superradiant clouds while only 20\% of them can survive the termination effect. This corresponds to a boson mass $\mu=7.6\times 10^{-13}\mathrm{eV}$ under the WR-BH spin model assumption described in the text.}
 \label{STSTcartoon}
\end{figure}

In this work, we make the first step towards a statistical test of superradiant clouds and their termination in realistic BH binaries. We start by evolving a realistic population of binary stars in our Galaxy into their stage of compact objects using the Stellar EVolution for N-body (SEVN) population synthesis code. Then focusing on the BH binaries, we adopt different spin models and compute the \change{rate $\Gamma_{nlm}$ at which the superradiant cloud grows} and the \change{correction} $\Delta\Gamma_{nlm}$ \change{to it} for each binary system. Systems with large enough \change{$\Gamma_{nlm}$ so that they are observationally relevant} are identified as \textit{participants} which potentially possess a boson cloud. Among the participants, systems with \change{effective superradiant growth rate} $\Gamma_{nlm}+\Delta\Gamma_{nlm}>0$ can survive termination and are identified as \textit{survivors}. The survival rate is then given by the ratio of the number of survivors and the number of participants (see Fig.~\ref{STSTcartoon} for illustration). Scanning over the unknown boson mass parameter $\mu$, we find that cloud modes with $l=m=1$ are robust in general while those with $l=m=2$ modes are vulnerable to superradiance termination. The survivors also tend to exhibit a higher superradiance rate, increasing their detectability.

\vskip 10pt

This paper is structured as follows. In Sec. \ref{Sec.Background}, we first briefly review the physics of a superradiant boson cloud near a rotating BH, with a focus on the termination effect from the binary. In Sec. \ref{Sec.Data}, we describe how we obtain a statistical sample of binary BHs by evolving the stellar binaries from their main sequence phase. We also introduce our spin models that describe how we estimate the BH spin based on the simulation output. Then in Sec.~\ref{Sec.STInfluence}, we examine the participants and survivors in the simulated samples and compute the survival rate for different cloud modes and boson masses. We conclude in Sec.~\ref{sec.Conclusion}.

\section{BH superradiance and its termination}
\label{Sec.Background}

In this section, we briefly review BH superradiance and the termination effect induced by a binary companion, while also laying out the relevant formulae for later analysis. Our discussions follow \citet{Baumann:2018vus}, \citet{Tong:2022bbl} and \citet{Fan:2023jjj}. See also \citet{Brito:2015oca} for a comprehensive review on BH superradiance.

\subsection{Superradiance and the gravitational atom}
\label{Sec.GA}

Consider a scalar field with mass $\mu$  bound by the gravity of a Kerr BH of mass $M$ and dimensionless spin $\tilde{a}$. The Klein-Gordon equation reads
\begin{align}
    \left(\square_{\rm Kerr}-\mu^2\right)\phi=0~.\label{KGeqn}
\end{align}
To make an analogy with the hydrogen atom, we introduce the gravitational fine structure constant $\alpha=M\mu$. In the case where $\alpha\lesssim1$, \eqref{KGeqn} admits a perturbative expansion in powers of $\alpha$ that also corresponds to a non-relativistic expansion.\footnote{While the Klein-Gordon equation can be solved numerically, reliable perturbative analytical solutions are also achievable when $\alpha<0.4$ \citep{Cannizzaro:2023jle}.} Rewriting $\phi=(2\mu)^{-1/2}e^{-i\mu t}\psi+\mathrm{c.c.}$, the Klein-Gordon equation reduces to a Schr\"{o}dinger-like equation
\begin{equation}
    i\,\partial_t\psi(t,\mathbf{x})=\left(-\frac{1}{2\mu}\partial^2_{\mathbf{x}}-\frac{\alpha}{r}+\mathcal{O}(\alpha^2)\right)\psi(t,\mathbf{x}) ~ . \label{SchrodingerEqFree}
\end{equation}
Focusing on the bound-state solutions, we impose the vanishing boundary condition at infinity and the in-going boundary condition at the outer horizon of the BH. This results in a set of bound states labeled by three integer quantum numbers $n,l,m$. The mode function reads
\begin{equation}
    \psi_{nlm}(r,\theta,\phi)\simeq R_{nl}(r)Y_{lm}(\theta,\phi)e^{-i(\omega_{nlm}-\mu)t} ~ , 
    \label{Eq.ModeFunction}
\end{equation}
where $R_{nl}(r)$ represents the radial wave function, and $Y_{lm}(\theta,\phi)$ the spherical harmonics. The corresponding eigenfrequency $\omega_{nlm}=E_{nlm}+i\Gamma_{nlm}$ is complex in general. The real part stands for the energy level of the mode \citep{Baumann:2018vus},
\begin{eqnarray}
     \nonumber   E_{nlm}&=&\mu\bigg[1-\frac{\alpha^2}{2 n^2}-\frac{\alpha^4}{8 n^4}-\frac{(3 n- l-1)\alpha^4}{n^4 (l+1/2)}\\&&+\frac{2 \tilde{a} m \alpha^5}{n^3 l (l+1/2) (l+1)}+\mathcal{O}(\alpha^6)\bigg] ~ .
\end{eqnarray}
The imaginary part, on the other hand, is inherent to the Kerr geometry and gives the superradiance rate. This becomes evident in (\ref{Eq.ModeFunction}), where we observe that the mode function is proportional to $e^{\Gamma_{nlm}t}$. Therefore, modes with $\Gamma_{nlm}>0$ are superradiant and growing, while those with $\Gamma_{nlm}<0$ are absorptive and decaying. The superradiance rate can be computed via the Detweiler approximation \citep{Detweiler:1980uk}\footnote{\change{A more accurate and general formula for the superradiant growth rate,  which also applies to the relativistic case can be found in \cite{Siemonsen:2022yyf}.}}, which gives
\begin{align}
   \hspace{-2mm} 
 \Gamma_{n00}&=-\frac{4}{n^3}\left(1+\sqrt{1-\tilde{a}^2}\right)\mu\alpha^5 ~ ,\label{00ModeAbsorptionRate} \\\hspace{-2mm} \Gamma_{nlm}&=2\tilde{r}_+\, C_{nl}g_{lm}(\tilde{a},\alpha,\omega_{nlm})(m\Omega_H-\omega_{nlm})\alpha^{4l+5} ~ ,
    \label{Eq.Imaginary}
\end{align}
where $\tilde{r}_+\equiv1+\sqrt{1-\tilde{a}^2}$ and $\Omega_H\equiv\tilde{a}/[2M(1+\sqrt{1-\tilde{a}^2})]$ is the angular velocity of the outer horizon. The coefficient functions are 
\begin{align}
    \hspace{-2mm}&C_{nl}\equiv\frac{2^{4l+1}(n+l)!}{n^{2l+4}(n-l-1)!}\,\left[\frac{l!}{(2l)!\,(2l+1!)}\right]^2 ~ , \\&\hspace{-2mm}g_{lm}(\tilde{a},\alpha,\omega)\equiv\prod_{k=1}^{l}\left(k^2\,(1-\tilde{a}^2)+(\tilde{a}m-2\tilde{r}_+ M\omega)^2\right) ~ . 
\end{align}

By angular momentum conservation, the growth of clouds with $m>0$ is accompanied by the decrease of the BH spin $\tilde{a}$. For isolated Kerr BHs, the superradiant growth of a given mode only ceases when the BH spin drops down to the saturation value where $m\Omega_H=\omega_{nlm}$. Subsequently, the cloud depletes solely due to the emission of GWs at a rate given by \citep{Brito:2014wla,Yoshino:2013ofa}
\begin{equation}
	\gamma_{nlm}=-B_{nl}\frac{S_{\text{c}}/m}{ M^2}\,\mu\alpha^{4l+10} ~ .\label{Eq.GWDepletion}
\end{equation}
Here, $S_{\text{c}}$ represents the total angular momentum of the cloud, and the numerical coefficients $B_{nl}$ can be found in \citep{Yoshino:2013ofa}. In this minimal scenario, each mode grows and depletes independently without explicit mutual interactions. In reality, however, both the environmental effects and the self-interaction of the boson field introduce the coupling between the cloud modes. In the next subsection, we discuss how the presence of a binary companion can lead to mode mixing and thus a significant termination effect.


\subsection{Effects of a binary companion and superradiance termination}

Now let us consider a superradiant BH within a binary system. In the presence of a companion with mass $M_*$, the Kerr geometry in \eqref{KGeqn} is deformed. To the leading order in the Newtonian approximation, this deformation is characterized by a tidal perturbation $V_*(t)$ in the gravitational potential. The Schr\"{o}dinger-like equation \eqref{SchrodingerEqFree} now becomes
\begin{equation}
    i\,\partial_t\psi(t,\mathbf{x})=\left(-\frac{1}{2\mu}\partial^2_{\mathbf{x}}-\frac{\alpha}{r}+V_*(t)+\mathcal{O}(\alpha^2)\right)\psi(t,\mathbf{x}) ~ . \label{SchrodingerEqWithVstar}
\end{equation}
\change{In Fermi normal coordinates, the Newtonian potential of the binary companion can be expanded using spherical harmonics $Y_{lm}(\theta,\phi)$ as follows 
\begin{equation}
    \begin{aligned}
        V_*&=-\alpha \, q \, \sum_{l_*\geq2}\sum_{|m_*|\leq l_*}\mathcal{E}_{l_*m_*}(\iota_*,\varphi_*)Y_{l_*m_*}(\theta,\phi)\\
        &\times\left(\frac{r^{l_*}}{R_*^{l_*+1}}\Theta(R_*-r)+\frac{R_*^{l_*}}{r^{l_*+1}}\Theta(r-R_*)\right) ~ ,  
    \end{aligned}\label{VstarMultipoleExpansion}
\end{equation}
with $q=M_*/M$ here denotes the mass ratio. $\Theta(x)$ is the Heaviside step function, and $\mathcal{E}$ is the tidal moment whose detailed definition can be found in \cite{Baumann:2019ztm}. }

By expanding both sides of the equation into eigenmodes \eqref{Eq.ModeFunction} in the free theory, we observe an off-diagonal contribution $\langle \psi_{n'l'm'}|V_*(t)| \psi_{nlm}\rangle$ that mixes different modes. Such mixing is typically weak and amounts to a small shift of eigenfrequencies $\omega_{nlm}\to \omega_{nlm}+\Delta \omega_{nlm}$ in the bound state spectrum. However, the shift in the imaginary direction ($\Delta \Gamma_{nlm}=\Im \Delta \omega_{nlm}$), despite being small, can be significant since the zeroth order contributions (see \eqref{Eq.Imaginary}) are also small. In particular, \citet{Tong:2022bbl} found that the mixing between slow superradiant modes and fast absorptive state can overturn a positive $\Gamma_{nlm}>0$ to a negative value, i.e. $\Gamma_{nlm}+\Delta \Gamma_{nlm}<0$, terminating superradiance. 


In detail, one can solve \eqref{SchrodingerEqWithVstar} and obtain the correction to the superradiance rate under the average co-rotation approximation \citep{Fan:2023jjj},
\begin{equation}
    \Delta \Gamma_{nlm}^{\rm (ACR)}\approx -\sum_{n'l'm'}\frac{(\Gamma_{nlm}-\Gamma_{n'l'm'})|\langle \psi_{n'l'm'}|V_*(t)| \psi_{nlm}\rangle|^2}{\left( E_{nlm}- E_{n'l'm'}\right)^2+\frac{M(1+q)}{p^3}(m-m')^2 }~ , 
    \label{Eq.DeltaGamma}
 \end{equation}
where we have summed over all possible $\psi_{n'l'm'}$ modes that can couple to $\psi_{nlm}$ according to the selection rules \citep{Baumann:2019ztm}
\begin{eqnarray}
    \nonumber &&-m'+m_*+m=0 ~ , \\ \nonumber &&l+l_*+l'=2k \quad \text{for }k\in \mathbb{Z} ~ , \\ && |l-l'|\leq l_*\leq l+l'~,~l_*\geq 2 ~.
    \label{Eq.SelectionRules}
\end{eqnarray}
In \eqref{Eq.DeltaGamma}, $p$ is the semi-latus rectum, which is related to the semi-major axis $a$ by $p\equiv a(1-e^2)$ with $e$ denoting the eccentricity.\footnote{We shall only deal with planar orbits in this paper.} The correction to the superradiance rate averaged over a period can be calculated by integrating over the true anomaly $\varphi_*$,
\begin{equation}
    \Delta\Gamma_{nlm}=\frac{1}{P}\int_0^{2\pi}\Delta\Gamma_{nlm}^{\rm (ACR)}\frac{{\rm d}\varphi_*}{\dot{\varphi_*}} ~ ,\label{STRateEq}
\end{equation}
where $P$ is the binary period.

\vskip 4pt

To have a sense of the typical time scale for relevant processes, we consider the case where the binary separation is large enough so that the second term in the denominator of (\ref{Eq.DeltaGamma}) becomes negligible. 
Let us examine the mode $\psi_{322}$ as an example. With a maximal BH spin $\tilde{a}=1$, the superradiance rate is 
\begin{eqnarray}
    \Gamma_{322}\simeq +5.\,{\rm Myr^{-1}}\left(\frac{\alpha}{0.1}\right)^{13}\left(\frac{M}{20M_{\odot}}\right)^{-1}~ , 
\end{eqnarray}
while the correction for the superradiance rate is 
\begin{eqnarray}
    \nonumber\Delta\Gamma_{322}\simeq&& -19.\,{\rm Myr^{-1}}\left(\frac{\alpha}{0.1}\right)^{-10} \left(\frac{M}{20M_{\odot}}\right)^{5}\\&&\times\left(\frac{q}{1.5}\right)^2\left(\frac{p}{5R_{\odot}}\right)^{-6}~,
\end{eqnarray}
with an eccentricity of $e=0.1$. Here $R_\odot\simeq 7.\times 10^8 {\rm m}$ denotes the solar radius. Notice that as the binary separation shrinks from infinity, $\Delta\Gamma$ quickly increases as $p^{-6}$, before being cut off by the second term in the denominator of \eqref{Eq.DeltaGamma} to $p^{-3}$. In comparison, the depletion rate of the cloud led by GW emission is much smaller,
\begin{align}
    \gamma_{322}\simeq -\,6.\times 10^{-10}\,{\rm Myr^{-1}}\left(\frac{\alpha}{0.1}\right)^{20}\left(\frac{M}{20M_{\odot}}\right)^{-1}~.
\end{align}
The GW emission rate of the BH binary itself is also small,
\begin{align}
    \nonumber \gamma_{\rm b}&\equiv\frac{1}{P}\frac{d P}{dt}\\
    &\simeq -\,0.1\,{\rm Myr^{-1}}\left(\frac{M}{20M_{\odot}}\right)^{3}\left(\frac{q}{1.5}\frac{1+q}{2.5}\right)\left(\frac{p}{5R_{\odot}}\right)^{-4}~,
\end{align}
which suggests that the orbital parameters $p,e$ can be approximated as constants when considering superradiance and its termination. In summary, the termination effect can easily dominate the evolution of the $\psi_{322}$ mode in this parameter regime. In the remaining part of the paper, we will apply \eqref{Eq.Imaginary} and \eqref{Eq.DeltaGamma} to simulate BH binaries and study how superradiance is affected by the termination effect.


\section{Simulating binary black holes in the Galaxy}
\label{Sec.Data}

To study how much a superradiant boson cloud near a BH can be affected by the presence of a companion in reality, we need to create a sample of simulated BH binaries with statistical properties similar to those in the Milky Way. In this section, we describe how we obtain this sample by using the Stellar EVolution for N-body (SEVN) population synthesis code \citep{spera2015mass,Spera:2017fyx,Spera:2018wnw}.

SEVN is able to reproduce the evolution of both single and binary stars from the initial zero-age main sequence (ZAMS) to a desired time of evolution that includes the compact object stages and the formation of BHs.\footnote{For a detailed list of evolutionary phases modeled in SEVN, see Table 1 in \citet{Spera:2018wnw})} This is by following a set of stellar evolution tracks that take into account many physical processes including those involved in binary systems \citep{Spera:2018wnw}. In particular in binary systems, when a large mass transfer occurs, the donor star will update the evolutionary track accordingly (this is a major difference from isolated systems that only follow fixed tracks). 
The reason why we prefer SEVN over other binary population-synthesis codes is that instead of using fitting formulas or integrating stellar evolution tracks at each run, it integrates the stellar evolution from a set of look-up tables, making it more flexible and less computationally expensive than previous models.

In this simulation, a total of $7\times10^7$ binary systems are evolved. The evolution starts from their initial stage of ZAMS where the two stars are still burning Hydrogen into Helium, and ends at their final stage when they both evolve into compact objects. The output data can be found \href{https://zenodo.org/records/13981305}{here} \footnote{\href{https://zenodo.org/records/13981305}{https://zenodo.org/records/13981305}}, and the python notebooks for generating the initial conditions and data analysis are available \href{https://github.com/jacquelynzhy/Statistical\_Superradiance}{here} \footnote{\href{https://github.com/jacquelynzhy/Statistical\_Superradiance}{https://github.com/jacquelynzhy/Statistical\_Superradiance}}.

\begin{figure*}[htbp]
	
	\begin{minipage}{0.34\linewidth}
		\vspace{3pt}
    \centerline{\includegraphics[width=\textwidth]{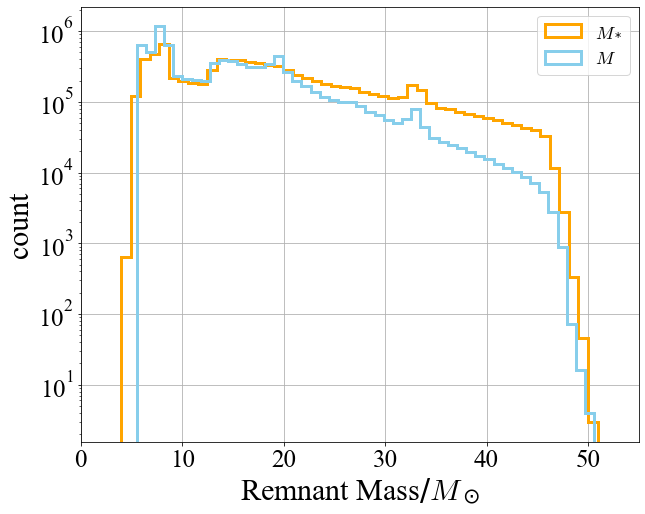}}
	\end{minipage}
	\begin{minipage}{0.34\linewidth}
		\vspace{3pt}
		\centerline{\includegraphics[width=\textwidth]{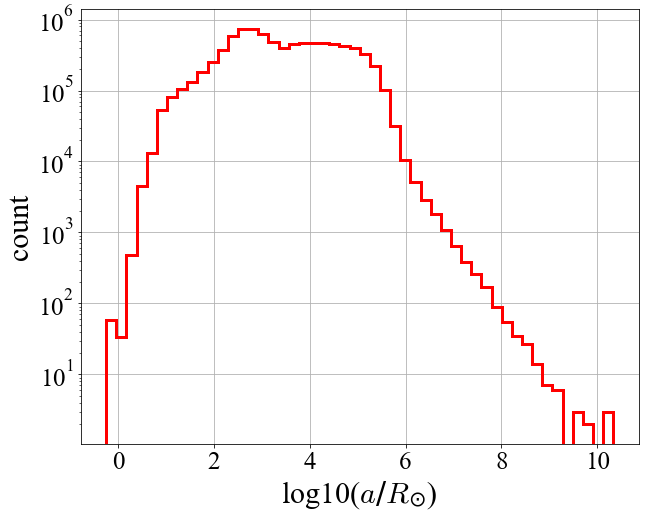}}
	 
	\end{minipage}
	\begin{minipage}{0.34\linewidth}
		\vspace{3pt}
		\centerline{\includegraphics[width=\textwidth]{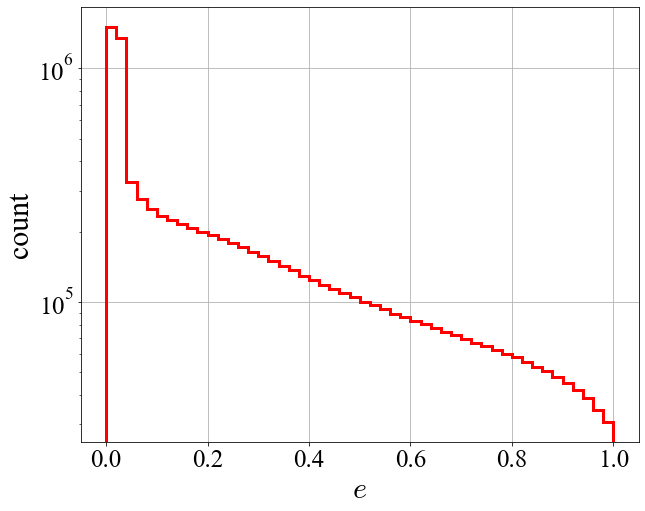}}
	 
	\end{minipage}
 
    \caption{The distribution of various binary BH parameters from the simulation output. \textit{Left panel:} The number count of remnant masses for the primary $M_*$ and secondary $M$. \textit{Middle panel:} The number count of semi-major axis $a$ for the remnant binary BH systems. \textit{Right panel:} The number count of eccentricity $e$.}
    \label{Fig.RemnParaCount}
\end{figure*}


\subsection{Initial condition}
\label{Sec.Ini}

In this subsection, we report the set of initial conditions that we have used to create our sample of compact object binaries. 
Except for the probability distribution for metallicities that requires special attention if we want to mimic the Galaxy, the rest of the conditions are almost identical to what has been used in the study from \citet{Spera:2018wnw}, which also agrees with the choice of \citet{Sana:2012px}.
\change{The justification of this set of parameters is extensively discussed in \cite{Spera:2018wnw}. Moreover, the effect of this particular choice on the resulting population of black hole binaries is the study of \cite{Iorio:2023sgz}. We shall comment more on this after we list the set of parameters adopted:}

\begin{itemize}

    \item The mass of the primary (larger mass) ZAMS star, denoted as $M_1$, follows a power-law distribution based on the Kroupa initial mass function \citep{Kroupa:2000iv}:
    \begin{equation}
       ~\quad \xi(M_1)\propto M_1^{-2.3} ~ , \quad M_1\in[10 M_{\odot},150M_{\odot}] ~ . 
    \end{equation}

    \item The mass of the secondary (smaller mass) ZAMS star $M_2$ is determined by the distribution of initial mass ratio $q_i$: 
    \begin{align}
        \nonumber ~  &\xi(q_i)\propto q_i^{-0.1} ~ , \\   & q_i=\frac{M_2}{M_1}\in[0.1,1]\quad \text{and}\quad M_2\geq10M_{\odot} ~ . 
    \end{align}

    \item The orbital properties considered are the period of the binaries $P_i$ and their eccentricities $e_i$ that are drawn from the following probability distributions:    \begin{eqnarray}
        &&\xi(\mathcal P_i)\propto \mathcal{P}_i^{-0.55} ~ , \quad \mathcal P_i\equiv\log_{10}\frac{P_i}{\rm day}\in[0.15,5.5]~ , \\&&\xi(e_i)\propto e_i^{-0.42} ~ , \quad  e_i\in[0,1] ~ . 
    \end{eqnarray}

    \item The original SEVN paper, \citet{Spera:2018wnw} chose different fixed values of metallicity for tests. However, since we aim to reproduce a sample of BH binaries that is representative of what can be found in a galaxy like the Milky Way, and the probability of forming a BH binary that coalesces within Hubble time is extremely dependent on the metallicity, with the merger efficiency that drops by three orders of magnitude from low to high metallicities \citep{Iorio:2023sgz}, instead of using a fixed metallicity, we adopt a distribution of metallicity in the range suggested by \citet{Spera:2018wnw}. Namely, we draw a sample of metallicities following the distribution reported in Figure 5 of \citet{lagarde2021deciphering}. In this study, they made use of the APOKASC catalogue that includes more than 6000 precise measurements of stellar ages \citep{pinsonneault:2018}. They then used the Besancon Galaxy Model \citep{robin:2003,lagarde:2017} to create a mock catalog where stars have different metallicities based on whether they belong to the thick or thin disk of the galaxy. In particular, for the thin disk, the metallicity is related to the age through $[{\rm Fe/H}] = -0.016\times \rm{age}+0.01$ while for the thick disk, two average values are used if the star belongs to the old or young disk ($[\rm{Fe}/H]=-0.8,-0.5$ dex respectively). With these assumptions and defining $\mathcal{Z}=[\rm{Fe}/H]$ as the metallicity, they obtain the distribution in their Figure 5 that we are able to fit with the sum of a Gaussian and a power-law function:
    \begin{eqnarray}
        \nonumber&&\xi(\mathcal{Z})=\xi_1(\mathcal{Z})+\xi_2(\mathcal{Z}) ~ ,~\text{with }\mathcal Z\in[0,0.5]\text{ and}\\
        \nonumber &&\xi_1(\mathcal Z)=0.03\times\mathcal Z^{0.45}+0.004  ~, \\&&\xi_2(\mathcal{Z})=0.11\times \exp\left[\frac{(\mathcal{Z}-0.61)^2}{2\times0.008^2}\right]~.
   \end{eqnarray}

    We have converted the Iron abundance metallicities in units of solar metallicities using $ Z=Z_{\odot}\times10^{\mathcal Z} $, with $Z_{\odot}=0.02$. 

    \item The dimensionless initial stellar spin \change{$\Omega$ is defined as the ratio between the angular velocity of the star $\Omega_S$ and $\Omega_C$, the critical angular velocity at which the star would be disrupted, \begin{align}\Omega=\frac{\Omega_S}{\Omega_C} \text{ with } \Omega_C = \sqrt{\frac{GM}{R^3_{\rm{eq}}}}~,
    \end{align}
    with $G$ the gravitational constant, $M$ the mass of the star and $R_{\rm{eq}}$ the equatorial radius. For $\Omega$, we adopt a uniform distribution from $0$ to maximal spin $1$}.     
\end{itemize}

In addition to these parameters, SEVN can also deal with different implementations of core-collapse supernovae. We have decided to be consistent with the conditions in \citet{Spera:2018wnw} and choose the rapid core-collapse model described in \citet{fryer2012compact}. With these conditions, we follow the entire stellar evolution from the ZAMS to the time when both objects become remnants.
\change{With this set of parameters we aim to reproduce a realistic population of black hole binaries for a galaxy similar to the Milky Way. Hence we warn the reader that the results found could be different in other types of galaxies with different morphologies. However, by changing the distribution of metallicities used in the code available online, the results can be adapted to any galaxy type. We also point out that we have tested the robustness of the model by varying the total number of binaries simulated. We found that the main conclusions of this work remain unchanged between a total binary number $\sim10^3$ and $\sim10^7$, which is what we used in this work. A thorough analysis of the variation of the parameters used in this work has been performed in \cite{Iorio:2023sgz} wh   ere they agree on the metallicity being the parameter that can affect the most the characteristics of the population of black hole binaries found. Based on the combined analysis of \cite{Iorio:2023sgz} and this work we can confidently state that there is no fine tuning in the choice of parameters and a slight variation would not affect our conclusion.}




\subsection{Binary BH properties}
\label{Sec.Output}

From the evolved sample of approximately $7\times10^7$ compact object binaries, we single out the binary BH systems\footnote{In principle, only one of the compact objects in the binary ought to be a BH. However, due to the astrophysical complexity of neutron stars and white dwarfs, for simplicity, we shall limit ourselves to the binary BH case in this work.} as our dataset for later tests of superradiance termination in Sec.~\ref{Sec.STInfluence}. This reduces the sample size to approximately $8\times10^6$. The distributions of primary and secondary masses, semi-major axes, and eccentricities are plotted in Fig.~\ref{Fig.RemnParaCount}. These plots show that the remnant BH masses vary between $5M_{\odot}$ and $50M_{\odot}$, while the semi-major axes 
fall within the range of around $10^0R_{\odot}$ to around $10^{10}R_{\odot}$. Smaller eccentricities (typically with $e < 0.1$) are also preferred by the statistics. 


\subsection{Spin models}

Despite the fact that SEVN can, in principle, account for the spin of the remnant BHs using several existing models, e.g. \citet{Belczynski:2017gds,Fuller:2019sxi,Bavera:2021evk}, the actual BH spins remain extremely uncertain. Intuitively, the spin of the BH in a binary system should be related to the binary separation, as the BH can acquire spin angular momentum through accretion and tidal spin-up processes, whose efficiency grows with shorter separation. Nevertheless, to date, there is no definitive answer for either the BH spin or its relation to the binary separation. However, superradiance termination is highly dependent on both the BH spin and the binary separation. Therefore, in order to be specific while also maintaining a certain level of generality, our strategy is to first work with a pre-existing Wolf-Rayet (WR) model studied by \citet{Bavera:2021evk,Zevin:2022wrw}, and then study our own plateau (PT) model that is inspired by the WR model but with adjustable parameters in the hope of incorporating the spin-model uncertainties. The analysis for superradiance termination is later performed for different choices of these parameters, with conclusions drawn solely from behaviours \textit{common} to the various choices.

\subsubsection{WR model}
\label{Sec.WRModel}

\begin{figure}[htbp]
    \centering
    \includegraphics[width=8cm]{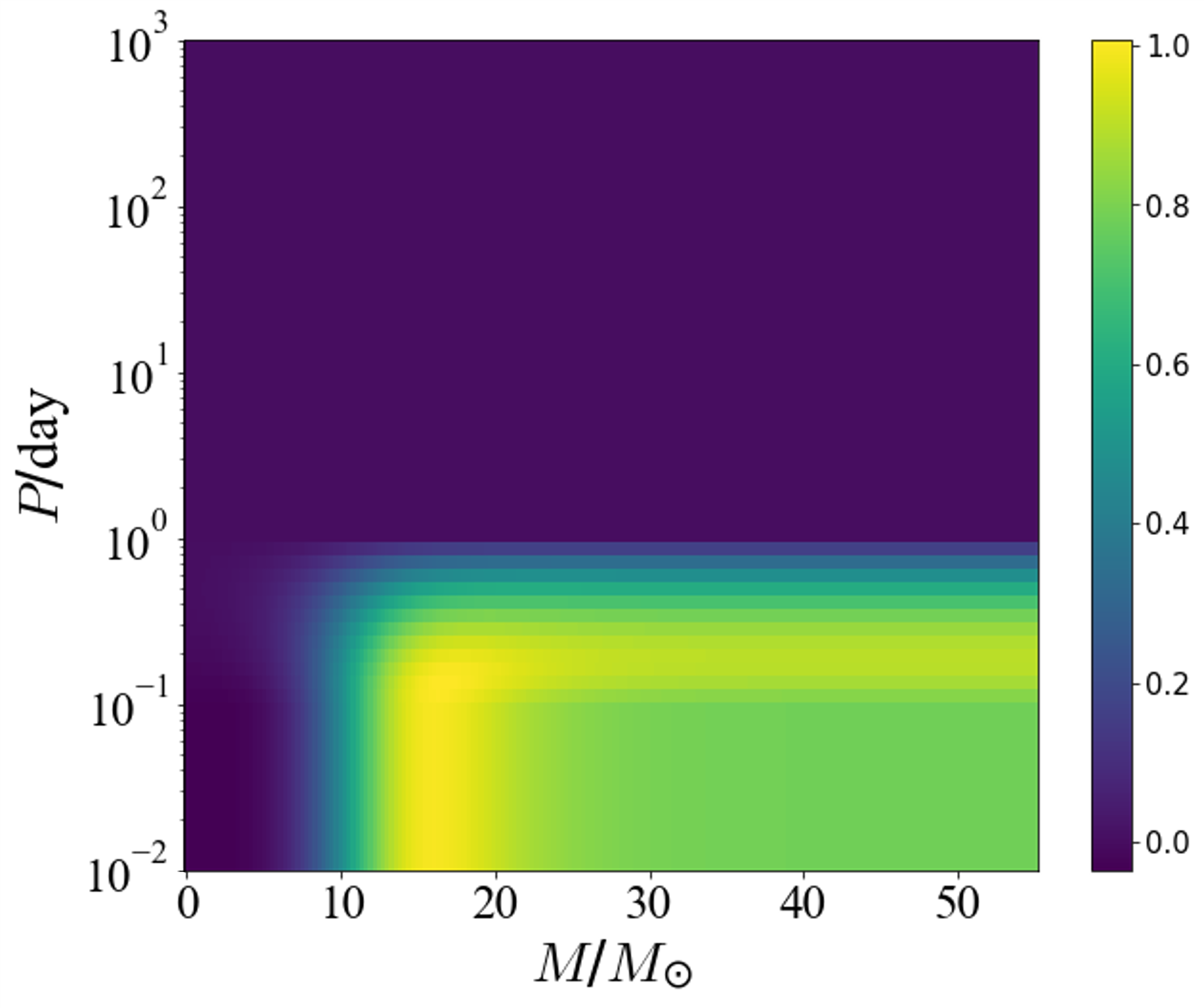}\\
    \caption{The BH spin dependence on the secondary BH mass and binary period in the WR spin model \eqref{WRModelDefEq}. The colour bar represents the magnitude of the BH spin $\tilde{a}$.}
    \label{Fig.WRModel}
\end{figure}

In the spin model of \citet{Zevin:2022wrw}, the progenitor of one of the BHs in the binary is a \change{Wolf-Rayet (WR)} star. 
\change{The reason for this choice is that WR stars are massive stars ($\gtrsim 20 M_{\odot}$) in a very late stage of their evolution and many studies consider them as the most likely progenitors of BH binaries (e.g. \citealt{Belczynski:2012,Higgins:2021,Iorio:2023sgz}).}
In such a system, the primary collapses into a BH which tends to have a low spin, and then the secondary loses its envelope, forming a WR star. Subsequently, tidal interaction spin-up the WR star, making it evolve into a fast-spinning BH \citep{Kushnir_2017, Qin:2018vaa, Bavera:2020inc, Olejak:2021iux, Fuller:2022ysb}. Since superradiance requires sufficient BH spin, we focus on the superradiance of the secondary BH which has a higher spin. Employing a semi-analytic fitting method, \citet{Bavera:2021evk} found that the spin of the secondary BH $\tilde{a}$ is well approximated by a quadratic polynomial of the logarithm of the orbital period $\log_{10} (\frac{P}{\text{day}})$:
\begin{equation}
    \hspace{-2mm}\tilde{a} = \left\{
    \begin{array}{ll}
    f^\alpha \log_{10}^2 \left(\frac{P}{\text{day}}\right) + f^\beta \log_{10} \left(\frac{P}{\text{day}}\right) &~,~  0.1 \leq \frac{P}{\text{day}} \leq 1 \\&\\
    0 &~,~  \frac{P}{\text{day}} > 1 \\&\\
    \tilde{a}|_{P=0.1\,{\rm day}} &~,~  \frac{P}{\text{day}} <0.1
    \end{array}
    \right.\label{WRModelDefEq}
\end{equation}
where $f^{(\alpha, \beta)} = -c_1^{(\alpha, \beta)} \left[c_2^{(\alpha, \beta)} + \exp\left(-c_3^{(\alpha, \beta)} M/{M_\odot}\right)\right]$, with coefficients $c_1^{(\alpha, \beta)}$, $c_2^{(\alpha, \beta)}$, and $c_3^{(\alpha, \beta)}$ determined from least-square regression (see \citet{Bavera:2021evk} for more details). The relation between the mass, spin of the secondary BH, and the binary period is shown in Fig.~\ref{Fig.WRModel}.

\subsubsection{PT model}
\label{Sec.PTModel}



	 

\begin{figure*}[htbp]

\centering

	\begin{minipage}{0.24\linewidth}
		\vspace{5pt}
\centerline{\includegraphics[width=4.8cm]{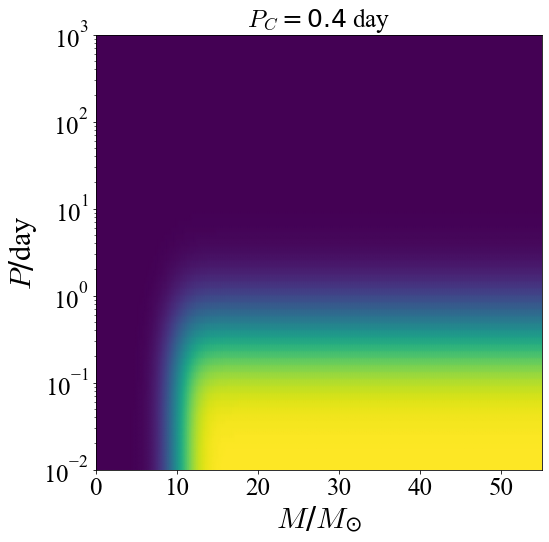}}
	\end{minipage}
	\begin{minipage}{0.24\linewidth}
		\vspace{5pt}
		\centerline{\includegraphics[width=4cm]{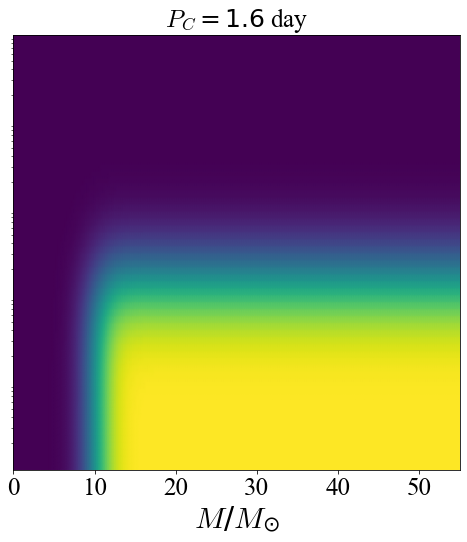}}
	 
	\end{minipage}
 	\begin{minipage}{0.24\linewidth}
		\vspace{5pt}
		\centerline{\includegraphics[width=4cm]{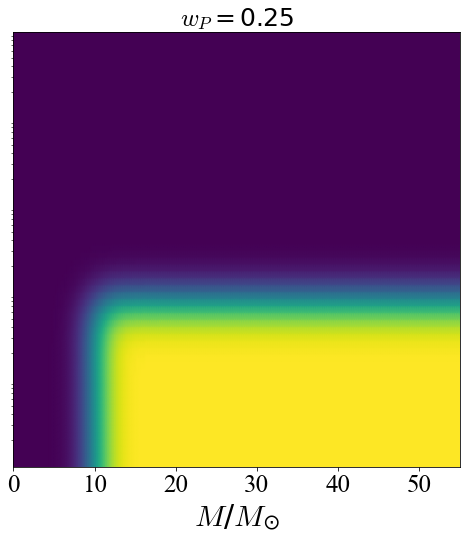}}
	 
	\end{minipage}
 	\begin{minipage}{0.24\linewidth}
		\vspace{5pt}
		\centerline{\includegraphics[width=4.7cm]{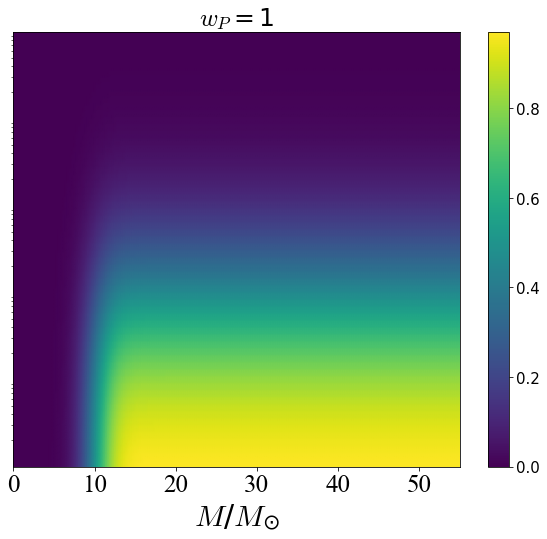}}
	 
	\end{minipage}
 \\
	\begin{minipage}{0.24\linewidth}
		\vspace{5pt}
\centerline{\includegraphics[width=4.8cm]{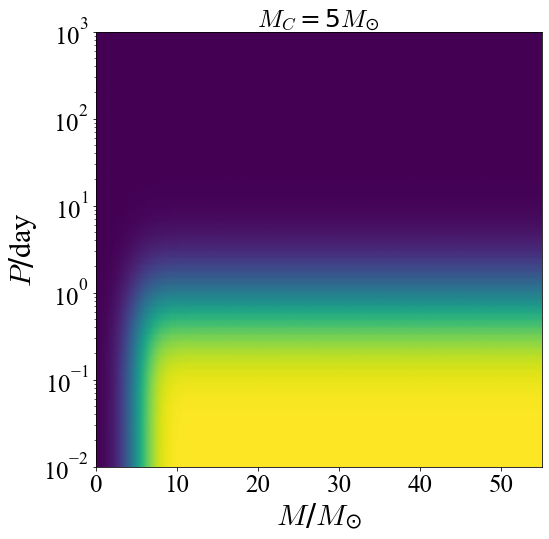}}
	\end{minipage}
	\begin{minipage}{0.24\linewidth}
		\vspace{5pt}
		\centerline{\includegraphics[width=4cm]{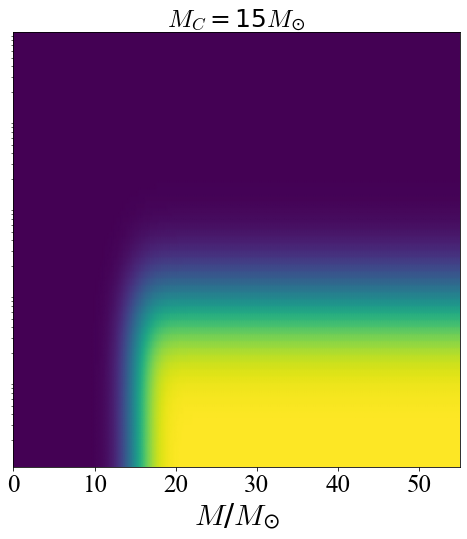}}
	 
	\end{minipage}
 	\begin{minipage}{0.24\linewidth}
		\vspace{5pt}
		\centerline{\includegraphics[width=4cm]{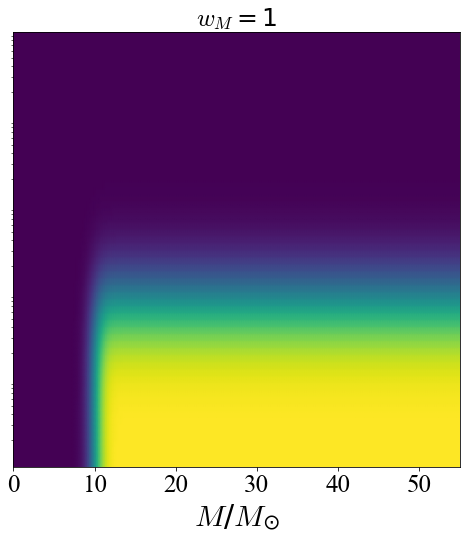}}
	 
	\end{minipage}
 	\begin{minipage}{0.24\linewidth}
		\vspace{5pt}
		\centerline{\includegraphics[width=4.7cm]{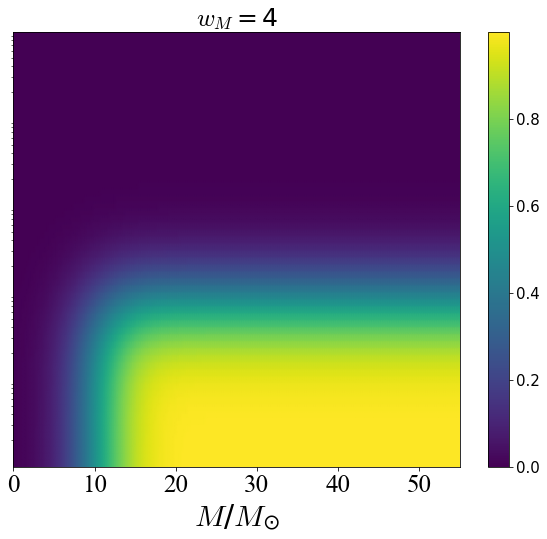}}
	 
	\end{minipage}
 
	\caption{The BH spin dependence on the secondary BH mass and binary period in the PT model \eqref{PTModelDefEq} with different parameter choices. Around the benchmark choice $P_C = 0.8$ day, $M_C = 10 M_{\odot},w_P = 0.5$, $w_M = 2$, we vary each parameter independently to display their functional dependence. In the \textit{upper left panel}, the period cutoff $P_C$ is adjusted and in the \textit{upper right panel}, the period width $w_P$ is adjusted. The \textit{lower left panel} corresponds to the adjusted mass cutoff $M_C$ and the \textit{lower right panel} corresponds to the adjusted mass width $w_M$. The colour bar represents the magnitude of the BH spin $\tilde{a}$.}
	\label{Fig.PTSpinModel}
\end{figure*}

The WR spin model is specifically designed for BHs that are evolved from WR star progenitors. In general, the BH spin is largely uncertain and hard to pin down with a specific model. To take these uncertainties into account and test the robustness of our prediction in more general scenarios, we choose to design our own spin model with different parameters, which are to be varied in the later analysis of superradiance termination. Consider
\begin{align}
    \tilde{a}=\frac{1}{4}\erfc\left(\frac{\ln P/P_C}{\sqrt{2}w_P}\right)\erfc\left(\frac{M_C-M}{\sqrt{2}w_M}\right) ~,\label{PTModelDefEq}
\end{align}
where $\erfc(x)$ is the complementary error function. This PT model is an emulation of the WR model with its essential features preserved. The PT model is characterized by a plateau region with two cutoffs ($P_C$ and $M_C$), along with two width parameters ($w_P$ and $w_M$) to adjust the sharpness of the cutoffs. To demonstrate the behaviour of the PT model, we plot \eqref{PTModelDefEq} for different parameter choices in Fig.~\ref{Fig.PTSpinModel}.

\vskip 10pt

\begin{figure}[htbp]
	\centering
	\includegraphics[width=8cm]{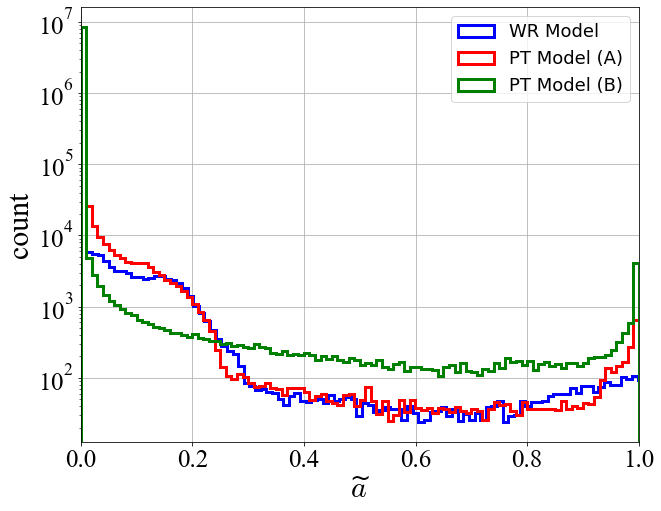}\\
	\caption{The spin distribution of secondary BHs in different spin models. For the PT model, parameter choice (\textcolor{red}{A}) corresponds to the benchmark model $P_C = 0.8$ day, $M_C = 10 M_{\odot},w_P = 0.5$, $w_M = 2$, while parameter choice (\textcolor{mygreen}{B}) corresponds to a variation of the model with $P_C = 0.4$ day, $M_C = 5 M_{\odot},w_P = 0.25$, $w_M = 1$. The BH masses and periods are taken from the output of the SEVN simulation, with the initial conditions stated in Sec.~\ref{Sec.Ini}. Despite the different predictions in the three cases, the spin distributions all peak around $\tilde{a}=0$ with a small fraction of BHs attaining large (even extremal) spins.}
 \label{Fig.PTSpin}
\end{figure}

\noindent \textbf{Spin distribution of the BHs} -- With the two spin models and the output from the SEVN, we obtain the spin distribution for the secondary BHs as depicted in Fig.~\ref{Fig.PTSpin}. For comparison, two sets of PT model parameters are chosen. Choice (A) corresponds to the benchmark values ($P_C = 0.8$ day, $M_C = 10 M_{\odot},w_P = 0.5$ and $w_M = 2$) while Choice (B) corresponds to a variation ($P_C = 0.4$ day, $M_C = 5 M_{\odot},w_P = 0.25$ and $w_M = 1$). From the plot, we observe that different spin models indeed give distinct predictions for the BH spin distribution. However, in all cases explored here, the spin distribution always peaks around $\tilde{a}=0$, while still having a small fraction of BHs with large (even extremal) spins.

\section{Statistical test of superradiance termination}
\label{Sec.STInfluence}

With the BH binaries sample simulated as described in the previous section, we are now in a position to study the statistics of superradiant BHs as well as the impact of the termination effect. In this section, we shall calculate the superradiance rate $\Gamma_{nlm}$ and the termination rate $\Delta \Gamma_{nlm}$ of the approximately $10^7$ BH binaries in the sample. The BH binaries in the sample are classified into participants and survivors according to these rates. Then we discuss their statistical correlations and move on to the survival probability of superradiant BHs for different boson masses. Due to their different predictions for BH spins, the discussions of the WR model and PT model are presented separately in the following subsections.


\subsection{WR model}

Employing the WR spin model introduced in Sec.~\ref{Sec.WRModel}, we can readily compute the superradiance rate using \eqref{Eq.Imaginary}. Depending on the BH mass and spin, the superradiant growth rate of boson clouds varies from binary to binary. Since we focus on superradiant BHs with observational significance, we select the BH binaries for which the superradiant growth time scale is less than \textit{(i).} the age of the universe, i.e. $0<\Gamma_{nlm}^{-1}<1.4\times 10^4$ Myr, and \textit{(ii).} the merger time scale of the binary, i.e. $0<\Gamma_{nlm}^{-1}<|\gamma_b|^{-1}$. This guarantees that the superradiant boson cloud can successfully grow within a reasonable amount of time without considering the termination effect. In addition, we impose \textit{(iii).} an upper bound on the gravitational fine structure constant $\alpha < 0.4$, due to the validity of the Detweiler approximation \eqref{Eq.Imaginary} \citep{Cannizzaro:2023jle}. This also ensures the validity of the non-relativistic approximation. BH binaries satisfying these three criteria are selected as \textit{participants}. On the other hand, the superradiance termination rate can also be computed using \eqref{STRateEq}. Among the participants, only those with $\Gamma_{nlm}+\Delta\Gamma_{nlm}>0$ can survive the termination effect and produce an abundantly populated boson cloud. Such BH binaries are selected as \textit{survivors}.

\begin{figure}[htbp]
	\centering
	
    \includegraphics[width=8cm]{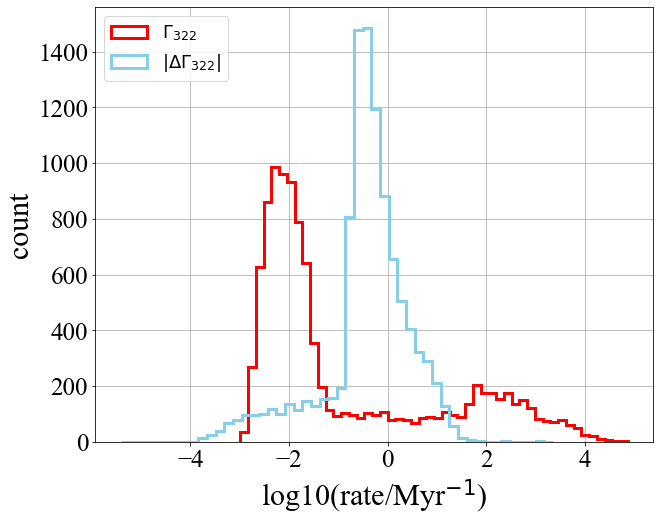}\\
 
    \includegraphics[width=8cm]{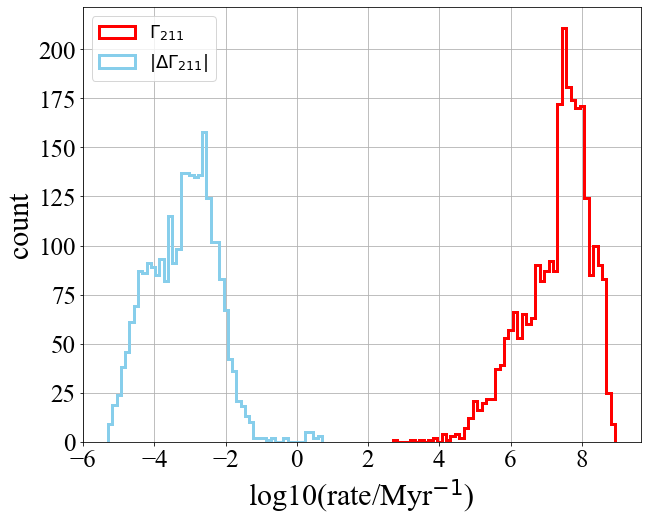}
    \caption{The distribution of the superradiance rate (\textcolor{myred}{red} line) and the absolute value of the termination rate (\textcolor{myblue}{blue} line) for the $\psi_{322}$ (\textit{upper panel}) and  $\psi_{211}$ (\textit{lower panel}) modes \change{of the participants}. Here we assumed the WR spin model and the boson mass is chosen to be $\mu=1.34\times10^{-12}$ eV. Here we see that the $\psi_{211}$ mode is robust against termination while roughly half of the $\psi_{322}$ modes are strongly affected.}
 \label{Fig.WRSTRate}
\end{figure}

\vskip 4pt

In Fig.~\ref{Fig.WRSTRate}, we plot the distribution of both the superradiance rate and termination rate \change{of the participants} for the leading two superradiant modes $\psi_{322}$ and $\psi_{211}$. We make two comments about this plot:
\begin{itemize}

    \item The first notable fact from Fig.~\ref{Fig.WRSTRate} is that the termination effect of the $\psi_{322}$ mode appears to be much stronger than that of the $\psi_{211}$ mode. Under the same parameter choice, $|\Delta\Gamma_{211}|$ is smaller than $\Gamma_{211}$ for \textit{all} participants while only roughly half of the participants have a $|\Delta\Gamma_{322}|$ smaller than $\Gamma_{322}$. This suggests all the $\psi_{211}$ participants survive the termination while only half of the $\psi_{322}$ participants do so. Such a distinction originates from the selection rules \eqref{Eq.SelectionRules}: via the $l_*=|m_*|=2$ tidal moment, the $l=m=2$ modes couple to the spherically symmetric $l=m=0$ modes, which possess the highest absorption rates (see \eqref{00ModeAbsorptionRate}). In contrast, the $l=m=1$ modes are protected from coupling to the $l=m=0$ modes by the selection rules \eqref{Eq.SelectionRules}. Instead, their termination effect is led by the $l=1$, $m=-1$ modes, whose absorption rates are considerably lower, making $l=m=1$ modes much more stable. Due to the pronounced stability of the $l=m=1$ modes, we shall only consider the $l=m=2$ modes for the survival probability henceforth.
    
    \item Another interesting aspect is that the total number of participants is smaller for $\psi_{211}$ than for $\psi_{322}$. This is natural because the superradiance threshold of $l=m=1$ modes is higher than that of the $l=m=2$ modes, which means $\psi_{211}$ requires a higher BH spin, hence forcing the participants to lie around the large-$\tilde{a}$ tail of the spin distribution in Fig.~\ref{Fig.PTSpin}. 
    
\end{itemize} 
Note that we have also checked by explicit calculation that the stability of $l=m=1$ persists for modes with higher principal quantum numbers with $n\geq 3$.
\vskip 4pt

\begin{figure}[htbp]
	\centering
	\includegraphics[width=8cm]{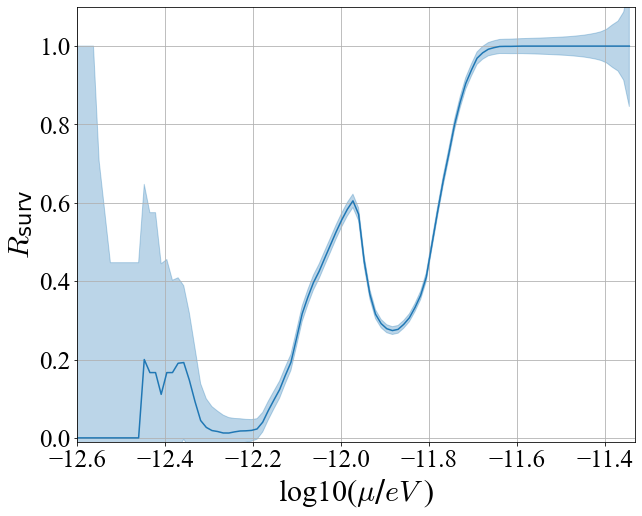}
	\caption{The survival rate $R_{\rm surv}$ as a function of boson mass parameter $\mu$ in the WR spin model. The shaded band indicates the uncertainty $\sigma (R_{\rm surv})$ estimated from \eqref{UncertainlyEstimatorDef}. Apparently the survival rate is overall an increasing function of the bosom mass, with less than $10\%$ of the superradiant participants eventually surviving the termination effect for small boson masses.}
 \label{Fig.WRRatio}
\end{figure}

To characterize the likelihood of a superradiant cloud surviving the termination from a binary companion, we define the survival rate as the ratio of the number of survivors and the number of participants
\begin{align}
    R_{\rm surv}\equiv\frac{N_{\rm surv}}{N_{\rm part}}~,
\end{align}
with $R_{\rm surv}\approx 0$ standing for complete cloud termination and $R_{\rm surv}\approx 1$ standing for assured survival.\footnote{$N_{\rm{part}}=0$ meaning no superradiance at all.} To reflect the statistical error in a sample of finite dimensionality, we also introduce an uncertainty estimator
\begin{align}
    \sigma (R_{\rm surv})\equiv\frac{1}{\sqrt{N_{\rm part}}}~,\label{UncertainlyEstimatorDef}
\end{align}
which is based on a Poisson distribution. We then scan the boson mass parameter $\mu$ and plot the survival rate of $\psi_{322}$ as a function of $\mu$ in Fig.~\ref{Fig.WRRatio}.\footnote{For $\psi_{211}$, we would have $R_{\rm surv}\simeq 1$.} The shaded band represents the uncertainty estimator \eqref{UncertainlyEstimatorDef}. From the plot, we observe that apart from the peak at around $\mu\sim10^{-12}$ eV (which will be discussed in more detail in the next subsection), the survival rate decreases for a smaller boson mass. In the large/small-$\mu$ limit, the survival rate asymptotes to one/zero, respectively. The uncertainty blows up at both ends due to the limited number of participants.\footnote{More technically, $N_{\rm part}$ drops to zero at the large-$\mu$ side because of the constraint (\textit{iii}) on $\alpha<0.4$, whereas at the small-$\mu$ side, it is due to the constraints (\textit{i}) and (\textit{ii}) on time scales.} Therefore, the termination effect can indeed dramatically influence the $\psi_{322}$ mode. In particular, for some boson masses, $\psi_{322}$ is \textit{confidently} terminated down to $R_{\rm surv}\pm\sigma (R_{\rm surv})<10\%$. We shall see later that these conclusions are robust against changing different spin models.

\subsection{PT model}
\label{PTresultSect}

\begin{figure}[htbp]
	\centering
	
    \includegraphics[width=8cm]{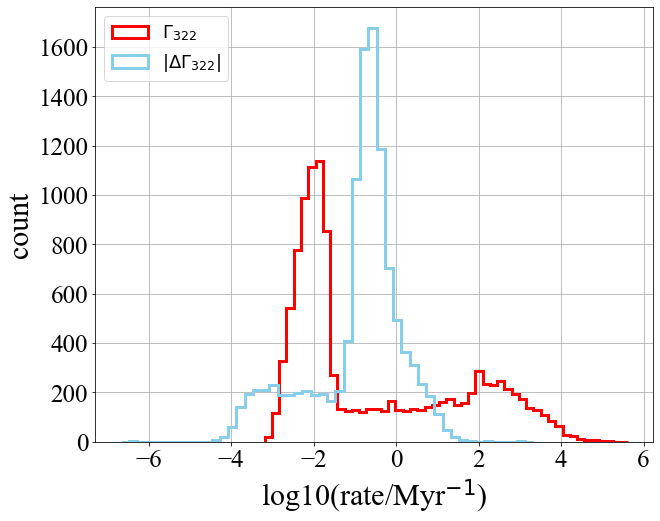}\\
 
    \includegraphics[width=8cm]{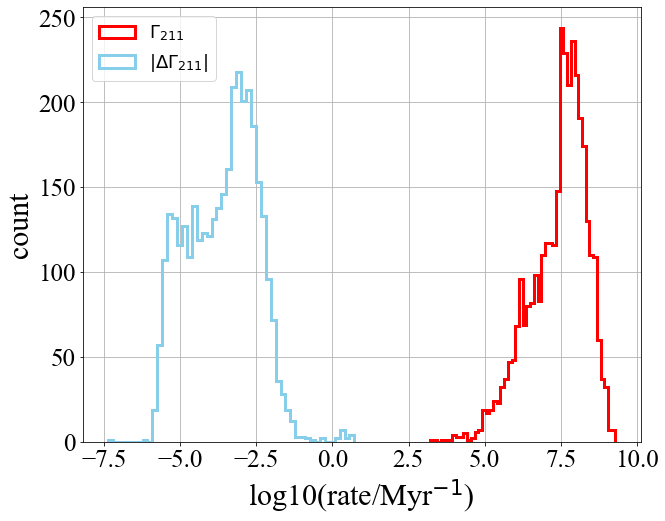}
    \caption{The distribution of the superradiance rate (the \textcolor{myred}{red} line) and the absolute value of the termination rate (the \textcolor{myblue}{blue} line) for the $\psi_{322}$ (\textit{upper panel}) and  $\psi_{211}$ (\textit{lower panel}) modes. Here we assumed the PT spin model with the benchmark choice ($P_C=0.8$ day, $w_P=0.5$, $M_C=10M_\odot$, $w_M=2$), and the boson mass is chosen to be $\mu=1.34\times10^{-12}$ eV. Again, the $\psi_{211}$ mode is robust against termination while roughly half of the $\psi_{322}$ modes are strongly affected.}
 \label{Fig.PTSTRate}
\end{figure}

\begin{figure*}[htbp]

\centering

	\begin{minipage}{0.45\linewidth}
		\vspace{5pt}
    \centerline{\includegraphics[width=8cm]{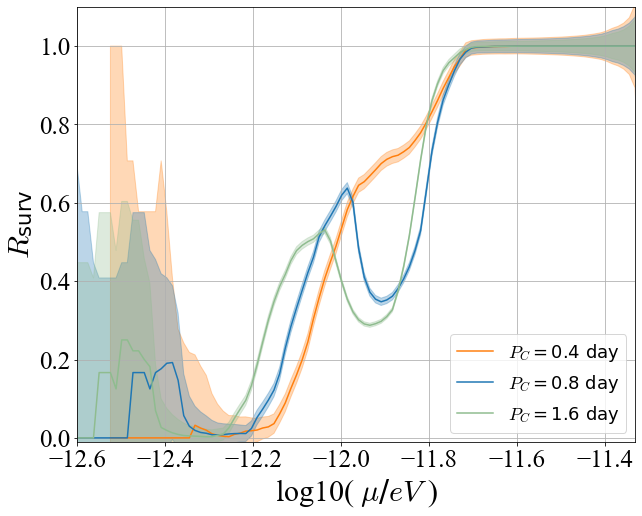}}
	\end{minipage}
	\begin{minipage}{0.45\linewidth}
		\vspace{5pt}
		\centerline{\includegraphics[width=8cm]{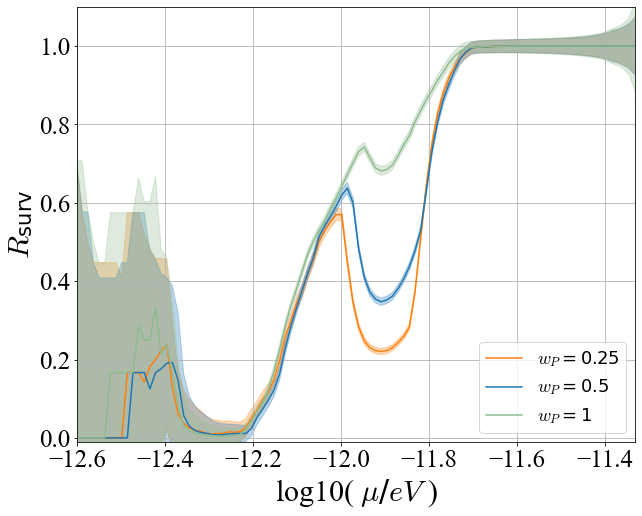}}
	 
	\end{minipage} \\
 	\begin{minipage}{0.45\linewidth}
		\vspace{5pt}
		\centerline{\includegraphics[width=8cm]{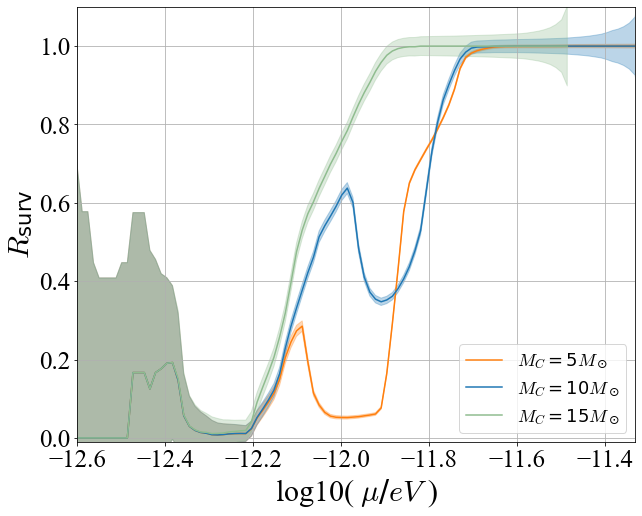}}

	\end{minipage} 
	\begin{minipage}{0.45\linewidth}
		\vspace{5pt}
		\centerline{\includegraphics[width=8cm]{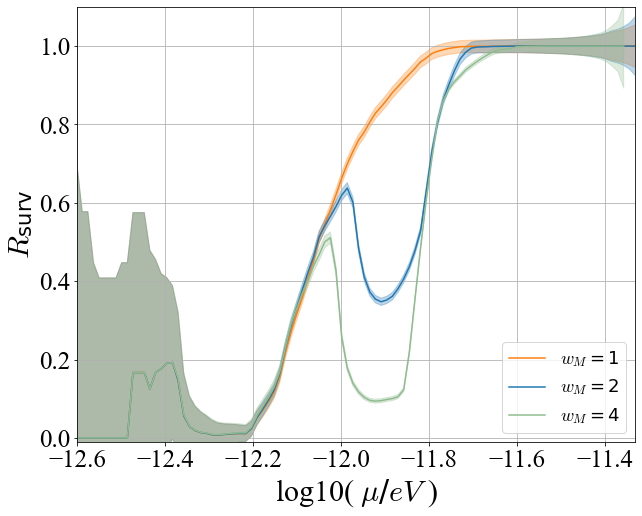}}
	 
	\end{minipage} 
 
    \caption{The survival rate in the PT spin model with varying parameters. The shaded band denotes the statistical uncertainties estimated using the Poisson noise \eqref{UncertainlyEstimatorDef}. Across all four panels, the \textcolor{myblue}{blue} lines correspond to the same benchmark parameter choice ($P_C = 0.8$ day, $M_C = 10 M_{\odot},w_P = 0.5$, $w_M = 2$). The \textit{upper left panel} shows variations in the period cutoff $P_C$, the \textit{upper right panel} shows variations in the period width $w_P$, the \textit{lower left panel} shows variations in the mass cutoff $M_C$, and the \textit{lower right panel} shows variations in the mass width $w_M$. These four panels are in one-to-one correspondence to the spin models in Fig.~\ref{Fig.PTSpinModel}.}\label{Fig.PTRate}
\end{figure*}

We now turn to the PT model and examine the survival rate of boson clouds in BH binaries with superradiance termination. The distribution of superradiance rate and termination rate is illustrated in Fig.~\ref{Fig.PTSTRate} with the benchmark choice $P_C=0.8$ days, $M_C=10M_\odot$, $w_P=0.5$ and $w_M$=2. It is again apparent that the $\psi_{211}$ mode is not at all influenced by the termination effect, henceforth we shall only focus on the mode $\psi_{322}$. To marginalize over the uncertainties in the spin models, we choose different combinations of the four parameters—$P_C$, $w_P$, $M_C$, and $w_M$ and observe their impact on the $R_{\rm surv}$-$\mu$ curve. For a fair comparison, we perturb around the benchmark parameter choice ($P_C = 0.8$ day, $M_C = 10 M_{\odot},w_P = 0.5$, $w_M = 2$) by varying one parameter at a time for all four parameters. The resulting survival rates as a function of the boson mass are plotted in Fig.~\ref{Fig.PTRate}. 
The shaded bands again represent the uncertainties \eqref{UncertainlyEstimatorDef} due to statistical fluctuations. We separate the discussions for each of the cases below:

\begin{itemize}
    \item The \textit{upper left panel} of Fig.~\ref{Fig.PTRate} illustrates how the $R_{\rm surv}$-$\mu$ curve depends on the period cutoff $P_C$. We notice that a lower $P_C$ corresponds to larger uncertainties. This occurs due to fewer BHs possessing adequate spin to sustain superradiance within a more stringent period cutoff. In a range of $\mu$ where the survival rate is still reliable, i.e. $R_{\rm surv}>\sigma(R_{\rm surv})$, the survival rate is lower for smaller $P_C$ when $\mu$ is small. One may understand it as a consequence of the fact that superradiant BHs in such cases typically appear inside close binaries, together with enhanced termination effect. For large boson masses, however, this trend is somehow unclear.

    \item The \textit{upper right panel} of Fig.~\ref{Fig.PTRate} displays the $R_{\rm surv}$-$\mu$ relation as we vary the period width parameter $w_P$. When $\mu$ is small, the differences are not significant. However, for a larger $\mu$, the survival rate is lower for a larger $w_P$, and the peak becomes less pronounced. This is because with a larger period width, more binaries with large separations are included. Since their termination effect is suppressed by the large separation, the survival rate consequently increases.

    \item The \textit{lower left panel} of Fig.~\ref{Fig.PTRate} gives the dependence on the mass cutoff $M_C$. From the plot, we observe that the peak is shifted towards smaller boson masses with a reduced survival rate. This occurs because selecting a smaller BH mass cutoff allows some small-mass black holes to exhibit significant spin and engage in superradiance, thus leading to smaller error bars. As $\mu$ increases, a substantial portion of larger mass black holes is excluded due to the non-relativistic limit $M\mu<0.4$. Consequently, smaller mass black holes can only engage in superradiance with a relatively lower superradiant rate, given the smaller $\alpha$, which facilitates an easier termination. Consequently, the survival rate at the peak decreases. This can also potentially explain the origin of the peak. From Fig.~\ref{Fig.RemnParaCount} we see that there are two peaks in the mass distribution lying at around $8M_\odot$ and $15\odot$. These two families of BH masses each corresponds to a preferred boson mass with different termination behaviour. The peak in the $R_{\rm surv}$-$\mu$ plots may thus arise from the competition between these two BH families. Consequently, lowering or raising the cutoff mass $M_C$ determines whether the $8M_\odot$-BH family can possess a large spin and enter as participants. Having $M_C=15M_\odot$ filters out the $8M_\odot$-BH family, leading to a monotonic behaviour of the survival rate dominated by the $15M_\odot$-BH family. 

    \item The \textit{lower right panel} of Fig.~\ref{Fig.PTRate} shows the survival rate for different mass widths $w_M$. A larger $w_M$ results in a less sharp mass cutoff, enabling lower mass black holes to potentially possess sufficient spin to sustain superradiance. As we claimed for the \textit{lower left panel}, this circumstance leads to an earlier peak appearance (at smaller $\mu$) with smaller $\mu$ values and reduces the survival rate. Additionally, it contributes to smaller uncertainties. On the other hand, tightening the mass width $w_M$ at $M_C=10M_\odot$ also filters out the $8M_\odot$-BH family, producing a monotonic survival rate curve.
\end{itemize}

Overall, we see that in agreement with the results in the WR model, the survival rate is approximately an increasing function of the boson mass and it asymptotes to zero/one when the boson mass is small/large. Depending on the spin models, the detailed shape of the survival rate curve can differ, but all of them drop below $R_{\rm surv}<10\%$ at $\mu\lesssim 0.5\times 10^{-12}$ eV. We also comment that the above behaviours persist for modes with higher principal quantum numbers ($\psi_{422},\psi_{522}$, etc.).

\subsection{Survivor statistics}

\begin{figure*}[htbp]
	\centering
	\includegraphics[width=0.9\textwidth]{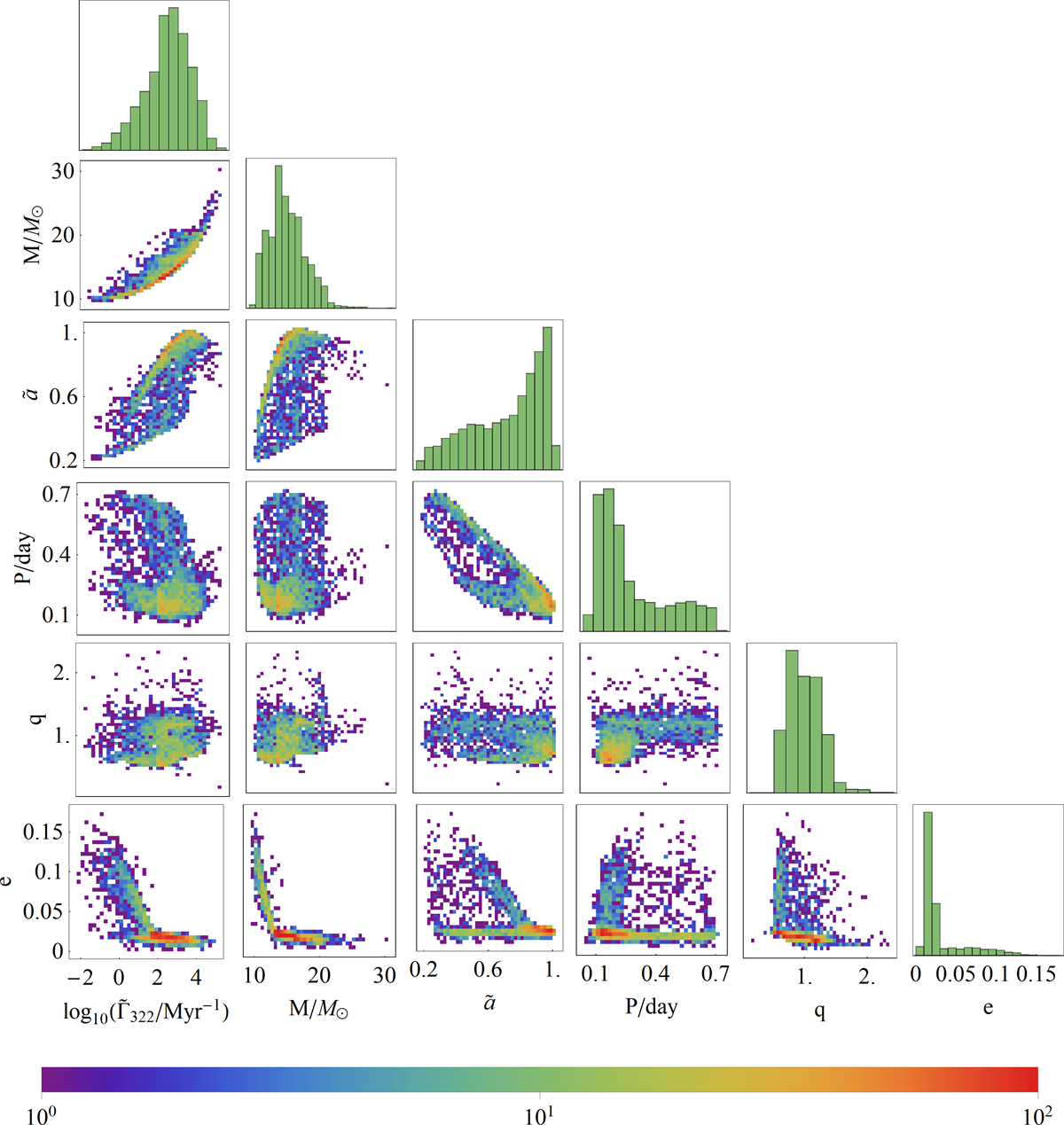}\\
	\caption{Density histograms for pairs of parameters of the survivors. Each subplot gives the distribution of survivors projected into a two-parameter subspace, with the colour indicating the counts in each bin. We have shown the correlation among the effective superradiance rate $\tilde\Gamma_{322}$, the BH mass $M$, the BH spin $\tilde{a}$, the binary period $P$, the mass ratio $q$ and the orbit eccentricity $e$. The diagonal corresponds to the marginal distribution for each parameter. Here the 2897 survivors are chosen from 10316 participants assuming the WR model with a boson mass $\mu=1.34\times 10^{-12}\mathrm{eV}$.}
 \label{Fig.SurvStat}
\end{figure*}

Assuming a spin model and a fixed boson mass parameter, we can obtain a list of survivors and study their statistics. In Fig.~\ref{Fig.SurvStat}, we plot the joint distributions of various parameters of the survivors including the effective superradiance rate $\tilde\Gamma_{322}\equiv\Gamma_{322}+\Delta \Gamma_{322}$, the BH mass $M$, the BH spin $\tilde{a}$, the binary period $P$, the mass ratio $q$ and the orbit eccentricity $e$. From comparing the distribution of the effective superradiance rate $\tilde\Gamma_{322}$ to that of the original superradiance rate $\Gamma_{322}$ of participants in Fig.~\ref{Fig.WRSTRate}, we see that the survivors are more superradiant on average. This can be read from their statistical means:
\begin{align}
    \left\langle\log_{10}(\tilde\Gamma_{322}/ \mathrm{Myr}^{-1})\right\rangle_{\rm surv}&\simeq 1.9~,\\
    \left\langle\log_{10}(\Gamma_{322} / \mathrm{Myr}^{-1})\right\rangle_{\rm part}&\simeq -0.8~.
\end{align}
Since $\Delta\Gamma_{322}<0$, these highly superradiant survivors also have a large positive $\Gamma_{322}$ to begin with. Therefore, boson clouds in highly superradiant BHs are more likely to withstand the tidal perturbation from a binary companion and survive the termination effect.\footnote{Hence the title.} From Fig.~\ref{Fig.SurvStat}, we see that a larger (effective) superradiant rate is positively correlated with the BH mass $M$, and hence to the gravitational fine structure constant $\alpha$. This suggests that survivors also tend to emit stronger GW signals (see e.g. \eqref{Eq.GWDepletion}) and are of more observational interest.

On the other hand, there does not seem to be a significant correlation between $P$ and $\tilde\Gamma_{322}$ since both superradiance and its termination intensify with a shorter period, where the former results from increased BH spin and the latter from increased tidal perturbations. There is a mild negative correlation between $e$ and $\tilde\Gamma_{322}$, suggesting a larger eccentricity may be correlated with a smaller BH mass and spin, thus also with a weaker superradiance rate. The strong correlation between $\tilde{a}$ and $M,P$ is, to a large extent, due to the WR spin model assumption.

\section{Conclusion and outlook}
\label{sec.Conclusion}
Superradiance termination can dramatically influence boson clouds in a BH binary, altering predictions about the ultralight boson itself. In this work, we provide the first examination of the impact of superradiance termination on boson clouds that potentially exist around BHs in a galaxy like the Milky Way. We simulated approximately $10^7$ BH binaries in the Galaxy using SEVN with realistic statistics. To obtain the BH spin, we adopt the spin model in a WR-BH binary as well as a plateau model with adjustable parameters to generalize the analysis. After computing the superradiance rate and the termination rate for each system in the dataset, we are able to determine the survival rate assuming different boson masses and spin models. Our analysis shows that the $l=m=1$ modes are generally resilient to the termination effect, thanks to the selection rule and a large mean binary separation. The $l=m=2$ modes, in contrast, can be significantly affected. For small boson masses (typically around $\mu\lesssim 0.5\times 10^{-12}$ eV), the survival rate can drop below $10\%$. We also find that BHs that survive the termination come from the highly superradiant tail of the participant sample, suggesting that observability is positively correlated with survivability. In summary, our work demonstrates the importance of superradiance termination in reality and paves the road for the statistical analysis of BH superradiance and the physics of ultralight bosons. 

\change{It is worth noting that in addition to the off-resonance termination effect that we considered in this work, the binary motion can trigger resonant transitions between the cloud modes \citep{Baumann:2019ztm}. Given the large binary separation in our dataset, such transitions, if exist at all, are typically of the hyperfine type and tend to decrease the angular quantum number, intensifying the termination process. Consequently, our analysis may underestimate the termination rate, and the survival rate we derived from neglecting resonant transitions can be seen as a conservative upper bound.}

There are certainly many questions left unanswered in the current work, and we hope to address them in the future. First of all, the great uncertainty in the BH spin models, although complicated, is surely something we hope to improve in the future. For instance, one important aspect that we have overlooked here is the interplay between mechanisms of BH spin generation and superradiance itself, as the very mechanisms that raise the BH spin might also be responsible for sourcing superradiance (and vice versa). We have also limited our study to binary BHs in this work, but in reality, the companion of a superradiant BH can also be a white dwarf or a neutron star. It is especially interesting to extend our statistical analysis to these cases since these binaries emit observable electromagnetic signals in addition to GWs. The termination of superradiance may come from sources other than a binary companion (e.g. accretion disks, self-interactions, etc.). Our cutoff on the fine structure constant also fails to account for the relativistic boson clouds, whose structure may be drastically different from the naive extrapolation. \change{Our calculation focuses on the off-resonance contribution to superradiance termination. Thus it would be nicer to include possible resonance contributions and investigate their significance in general.} It is therefore crucial to include all these aspects in future works for a more complete statistical analysis of superradiance.



\section*{Data Availability}
\change{All the data and results used in this work are publicly available at \url{https://zenodo.org/records/13981305} along with the Python code we used for analysing the data, \url{https://github.com/jacquelynzhy/Statistical Superradiance}. Please contact the corresponding author for further assistance in using the code. }

\section*{Acknowledgment} 

We would like to thank Giuliano Iorio for helping us understand SEVN, and Elisa Bortolas for initial collaboration. We would also like to thank Cosimo Bambi, Leo W.H. Fung and Yi Wang for fruitful discussions. This work was supported in part by the National Key R\&D Program of China (2021YFC2203100),  CRF grant C6017-20GF and GRF grant 16306422 by the RGC of Hong Kong SAR. XT is supported by STFC consolidated grants ST/T000694/1 and ST/X000664/1. GM is supported by the Collaborative Research Fund under Grant No. C6017-20G which is issued by the Research Grants Council of Hong Kong S.A.R.


\bibliography{STSTRefs}

\begin{thebibliography}{74}%
\makeatletter
\providecommand \@ifxundefined [1]{%
 \@ifx{#1\undefined}
}%
\providecommand \@ifnum [1]{%
 \ifnum #1\expandafter \@firstoftwo
 \else \expandafter \@secondoftwo
 \fi
}%
\providecommand \@ifx [1]{%
 \ifx #1\expandafter \@firstoftwo
 \else \expandafter \@secondoftwo
 \fi
}%
\providecommand \natexlab [1]{#1}%
\providecommand \enquote  [1]{``#1''}%
\providecommand \bibnamefont  [1]{#1}%
\providecommand \bibfnamefont [1]{#1}%
\providecommand \citenamefont [1]{#1}%
\providecommand \href@noop [0]{\@secondoftwo}%
\providecommand \href [0]{\begingroup \@sanitize@url \@href}%
\providecommand \@href[1]{\@@startlink{#1}\@@href}%
\providecommand \@@href[1]{\endgroup#1\@@endlink}%
\providecommand \@sanitize@url [0]{\catcode `\\12\catcode `\$12\catcode
  `\&12\catcode `\#12\catcode `\^12\catcode `\_12\catcode `\%12\relax}%
\providecommand \@@startlink[1]{}%
\providecommand \@@endlink[0]{}%
\providecommand \url  [0]{\begingroup\@sanitize@url \@url }%
\providecommand \@url [1]{\endgroup\@href {#1}{\urlprefix }}%
\providecommand \urlprefix  [0]{URL }%
\providecommand \Eprint [0]{\href }%
\providecommand \doibase [0]{http://dx.doi.org/}%
\providecommand \selectlanguage [0]{\@gobble}%
\providecommand \bibinfo  [0]{\@secondoftwo}%
\providecommand \bibfield  [0]{\@secondoftwo}%
\providecommand \translation [1]{[#1]}%
\providecommand \BibitemOpen [0]{}%
\providecommand \bibitemStop [0]{}%
\providecommand \bibitemNoStop [0]{.\EOS\space}%
\providecommand \EOS [0]{\spacefactor3000\relax}%
\providecommand \BibitemShut  [1]{\csname bibitem#1\endcsname}%
\let\auto@bib@innerbib\@empty
\bibitem [{\citenamefont {Peccei}\ and\ \citenamefont
  {Quinn}(1977)}]{Peccei:1977hh}%
  \BibitemOpen
  \bibfield  {author} {\bibinfo {author} {\bibfnamefont {R.~D.}\ \bibnamefont
  {Peccei}}\ and\ \bibinfo {author} {\bibfnamefont {H.~R.}\ \bibnamefont
  {Quinn}},\ }\href {\doibase 10.1103/PhysRevLett.38.1440} {\bibfield
  {journal} {\bibinfo  {journal} {Phys. Rev. Lett.}\ }\textbf {\bibinfo
  {volume} {38}},\ \bibinfo {pages} {1440} (\bibinfo {year}
  {1977})}\BibitemShut {NoStop}%
\bibitem [{\citenamefont {Weinberg}(1978)}]{Weinberg:1977ma}%
  \BibitemOpen
  \bibfield  {author} {\bibinfo {author} {\bibfnamefont {S.}~\bibnamefont
  {Weinberg}},\ }\href {\doibase 10.1103/PhysRevLett.40.223} {\bibfield
  {journal} {\bibinfo  {journal} {Phys. Rev. Lett.}\ }\textbf {\bibinfo
  {volume} {40}},\ \bibinfo {pages} {223} (\bibinfo {year} {1978})}\BibitemShut
  {NoStop}%
\bibitem [{\citenamefont {Wilczek}(1978)}]{Wilczek:1977pj}%
  \BibitemOpen
  \bibfield  {author} {\bibinfo {author} {\bibfnamefont {F.}~\bibnamefont
  {Wilczek}},\ }\href {\doibase 10.1103/PhysRevLett.40.279} {\bibfield
  {journal} {\bibinfo  {journal} {Phys. Rev. Lett.}\ }\textbf {\bibinfo
  {volume} {40}},\ \bibinfo {pages} {279} (\bibinfo {year} {1978})}\BibitemShut
  {NoStop}%
\bibitem [{\citenamefont {Kim}(1979)}]{Kim:1979if}%
  \BibitemOpen
  \bibfield  {author} {\bibinfo {author} {\bibfnamefont {J.~E.}\ \bibnamefont
  {Kim}},\ }\href {\doibase 10.1103/PhysRevLett.43.103} {\bibfield  {journal}
  {\bibinfo  {journal} {Phys. Rev. Lett.}\ }\textbf {\bibinfo {volume} {43}},\
  \bibinfo {pages} {103} (\bibinfo {year} {1979})}\BibitemShut {NoStop}%
\bibitem [{\citenamefont {Svrcek}\ and\ \citenamefont
  {Witten}(2006)}]{Svrcek:2006yi}%
  \BibitemOpen
  \bibfield  {author} {\bibinfo {author} {\bibfnamefont {P.}~\bibnamefont
  {Svrcek}}\ and\ \bibinfo {author} {\bibfnamefont {E.}~\bibnamefont
  {Witten}},\ }\href {\doibase 10.1088/1126-6708/2006/06/051} {\bibfield
  {journal} {\bibinfo  {journal} {JHEP}\ }\textbf {\bibinfo {volume} {06}},\
  \bibinfo {pages} {051} (\bibinfo {year} {2006})},\ \Eprint
  {http://arxiv.org/abs/hep-th/0605206} {arXiv:hep-th/0605206} \BibitemShut
  {NoStop}%
\bibitem [{\citenamefont {Arvanitaki}\ \emph {et~al.}(2010)\citenamefont
  {Arvanitaki}, \citenamefont {Dimopoulos}, \citenamefont {Dubovsky},
  \citenamefont {Kaloper},\ and\ \citenamefont
  {March-Russell}}]{Arvanitaki:2009fg}%
  \BibitemOpen
  \bibfield  {author} {\bibinfo {author} {\bibfnamefont {A.}~\bibnamefont
  {Arvanitaki}}, \bibinfo {author} {\bibfnamefont {S.}~\bibnamefont
  {Dimopoulos}}, \bibinfo {author} {\bibfnamefont {S.}~\bibnamefont
  {Dubovsky}}, \bibinfo {author} {\bibfnamefont {N.}~\bibnamefont {Kaloper}}, \
  and\ \bibinfo {author} {\bibfnamefont {J.}~\bibnamefont {March-Russell}},\
  }\href {\doibase 10.1103/PhysRevD.81.123530} {\bibfield  {journal} {\bibinfo
  {journal} {Phys. Rev. D}\ }\textbf {\bibinfo {volume} {81}},\ \bibinfo
  {pages} {123530} (\bibinfo {year} {2010})},\ \Eprint
  {http://arxiv.org/abs/0905.4720} {arXiv:0905.4720 [hep-th]} \BibitemShut
  {NoStop}%
\bibitem [{\citenamefont {Mehta}\ \emph {et~al.}(2021)\citenamefont {Mehta},
  \citenamefont {Demirtas}, \citenamefont {Long}, \citenamefont {Marsh},
  \citenamefont {McAllister},\ and\ \citenamefont {Stott}}]{Mehta:2021pwf}%
  \BibitemOpen
  \bibfield  {author} {\bibinfo {author} {\bibfnamefont {V.~M.}\ \bibnamefont
  {Mehta}}, \bibinfo {author} {\bibfnamefont {M.}~\bibnamefont {Demirtas}},
  \bibinfo {author} {\bibfnamefont {C.}~\bibnamefont {Long}}, \bibinfo {author}
  {\bibfnamefont {D.~J.~E.}\ \bibnamefont {Marsh}}, \bibinfo {author}
  {\bibfnamefont {L.}~\bibnamefont {McAllister}}, \ and\ \bibinfo {author}
  {\bibfnamefont {M.~J.}\ \bibnamefont {Stott}},\ }\href {\doibase
  10.1088/1475-7516/2021/07/033} {\bibfield  {journal} {\bibinfo  {journal}
  {JCAP}\ }\textbf {\bibinfo {volume} {07}},\ \bibinfo {pages} {033} (\bibinfo
  {year} {2021})},\ \Eprint {http://arxiv.org/abs/2103.06812} {arXiv:2103.06812
  [hep-th]} \BibitemShut {NoStop}%
\bibitem [{\citenamefont {Dine}\ and\ \citenamefont
  {Fischler}(1983)}]{Dine:1982ah}%
  \BibitemOpen
  \bibfield  {author} {\bibinfo {author} {\bibfnamefont {M.}~\bibnamefont
  {Dine}}\ and\ \bibinfo {author} {\bibfnamefont {W.}~\bibnamefont
  {Fischler}},\ }\href {\doibase 10.1016/0370-2693(83)90639-1} {\bibfield
  {journal} {\bibinfo  {journal} {Phys. Lett. B}\ }\textbf {\bibinfo {volume}
  {120}},\ \bibinfo {pages} {137} (\bibinfo {year} {1983})}\BibitemShut
  {NoStop}%
\bibitem [{\citenamefont {Preskill}\ \emph {et~al.}(1983)\citenamefont
  {Preskill}, \citenamefont {Wise},\ and\ \citenamefont
  {Wilczek}}]{Preskill:1982cy}%
  \BibitemOpen
  \bibfield  {author} {\bibinfo {author} {\bibfnamefont {J.}~\bibnamefont
  {Preskill}}, \bibinfo {author} {\bibfnamefont {M.~B.}\ \bibnamefont {Wise}},
  \ and\ \bibinfo {author} {\bibfnamefont {F.}~\bibnamefont {Wilczek}},\ }\href
  {\doibase 10.1016/0370-2693(83)90637-8} {\bibfield  {journal} {\bibinfo
  {journal} {Phys. Lett. B}\ }\textbf {\bibinfo {volume} {120}},\ \bibinfo
  {pages} {127} (\bibinfo {year} {1983})}\BibitemShut {NoStop}%
\bibitem [{\citenamefont {Abbott}\ and\ \citenamefont
  {Sikivie}(1983)}]{Abbott:1982af}%
  \BibitemOpen
  \bibfield  {author} {\bibinfo {author} {\bibfnamefont {L.~F.}\ \bibnamefont
  {Abbott}}\ and\ \bibinfo {author} {\bibfnamefont {P.}~\bibnamefont
  {Sikivie}},\ }\href {\doibase 10.1016/0370-2693(83)90638-X} {\bibfield
  {journal} {\bibinfo  {journal} {Phys. Lett. B}\ }\textbf {\bibinfo {volume}
  {120}},\ \bibinfo {pages} {133} (\bibinfo {year} {1983})}\BibitemShut
  {NoStop}%
\bibitem [{\citenamefont {Hui}\ \emph {et~al.}(2017)\citenamefont {Hui},
  \citenamefont {Ostriker}, \citenamefont {Tremaine},\ and\ \citenamefont
  {Witten}}]{Hui:2016ltb}%
  \BibitemOpen
  \bibfield  {author} {\bibinfo {author} {\bibfnamefont {L.}~\bibnamefont
  {Hui}}, \bibinfo {author} {\bibfnamefont {J.~P.}\ \bibnamefont {Ostriker}},
  \bibinfo {author} {\bibfnamefont {S.}~\bibnamefont {Tremaine}}, \ and\
  \bibinfo {author} {\bibfnamefont {E.}~\bibnamefont {Witten}},\ }\href
  {\doibase 10.1103/PhysRevD.95.043541} {\bibfield  {journal} {\bibinfo
  {journal} {Phys. Rev. D}\ }\textbf {\bibinfo {volume} {95}},\ \bibinfo
  {pages} {043541} (\bibinfo {year} {2017})},\ \Eprint
  {http://arxiv.org/abs/1610.08297} {arXiv:1610.08297 [astro-ph.CO]}
  \BibitemShut {NoStop}%
\bibitem [{\citenamefont {Detweiler}(1980)}]{Detweiler:1980uk}%
  \BibitemOpen
  \bibfield  {author} {\bibinfo {author} {\bibfnamefont {S.~L.}\ \bibnamefont
  {Detweiler}},\ }\href {\doibase 10.1103/PhysRevD.22.2323} {\bibfield
  {journal} {\bibinfo  {journal} {Phys. Rev. D}\ }\textbf {\bibinfo {volume}
  {22}},\ \bibinfo {pages} {2323} (\bibinfo {year} {1980})}\BibitemShut
  {NoStop}%
\bibitem [{\citenamefont {Damour}\ \emph {et~al.}(1976)\citenamefont {Damour},
  \citenamefont {Deruelle},\ and\ \citenamefont {Ruffini}}]{Damour:1976kh}%
  \BibitemOpen
  \bibfield  {author} {\bibinfo {author} {\bibfnamefont {T.}~\bibnamefont
  {Damour}}, \bibinfo {author} {\bibfnamefont {N.}~\bibnamefont {Deruelle}}, \
  and\ \bibinfo {author} {\bibfnamefont {R.}~\bibnamefont {Ruffini}},\ }\href
  {\doibase 10.1007/BF02725534} {\bibfield  {journal} {\bibinfo  {journal}
  {Lett. Nuovo Cim.}\ }\textbf {\bibinfo {volume} {15}},\ \bibinfo {pages}
  {257} (\bibinfo {year} {1976})}\BibitemShut {NoStop}%
\bibitem [{\citenamefont {Brito}\ \emph
  {et~al.}(2015{\natexlab{a}})\citenamefont {Brito}, \citenamefont {Cardoso},\
  and\ \citenamefont {Pani}}]{Brito:2015oca}%
  \BibitemOpen
  \bibfield  {author} {\bibinfo {author} {\bibfnamefont {R.}~\bibnamefont
  {Brito}}, \bibinfo {author} {\bibfnamefont {V.}~\bibnamefont {Cardoso}}, \
  and\ \bibinfo {author} {\bibfnamefont {P.}~\bibnamefont {Pani}},\ }\href
  {\doibase 10.1007/978-3-319-19000-6} {\bibfield  {journal} {\bibinfo
  {journal} {Lect. Notes Phys.}\ }\textbf {\bibinfo {volume} {906}},\ \bibinfo
  {pages} {pp.1} (\bibinfo {year} {2015}{\natexlab{a}})},\ \Eprint
  {http://arxiv.org/abs/1501.06570} {arXiv:1501.06570 [gr-qc]} \BibitemShut
  {NoStop}%
\bibitem [{\citenamefont {Cannizzaro}\ \emph {et~al.}(2024)\citenamefont
  {Cannizzaro}, \citenamefont {Sberna}, \citenamefont {Green},\ and\
  \citenamefont {Hollands}}]{Cannizzaro:2023jle}%
  \BibitemOpen
  \bibfield  {author} {\bibinfo {author} {\bibfnamefont {E.}~\bibnamefont
  {Cannizzaro}}, \bibinfo {author} {\bibfnamefont {L.}~\bibnamefont {Sberna}},
  \bibinfo {author} {\bibfnamefont {S.~R.}\ \bibnamefont {Green}}, \ and\
  \bibinfo {author} {\bibfnamefont {S.}~\bibnamefont {Hollands}},\ }\href
  {\doibase 10.1103/PhysRevLett.132.051401} {\bibfield  {journal} {\bibinfo
  {journal} {Phys. Rev. Lett.}\ }\textbf {\bibinfo {volume} {132}},\ \bibinfo
  {pages} {051401} (\bibinfo {year} {2024})},\ \Eprint
  {http://arxiv.org/abs/2309.10021} {arXiv:2309.10021 [gr-qc]} \BibitemShut
  {NoStop}%
\bibitem [{\citenamefont {Arvanitaki}\ and\ \citenamefont
  {Dubovsky}(2011)}]{Arvanitaki:2010sy}%
  \BibitemOpen
  \bibfield  {author} {\bibinfo {author} {\bibfnamefont {A.}~\bibnamefont
  {Arvanitaki}}\ and\ \bibinfo {author} {\bibfnamefont {S.}~\bibnamefont
  {Dubovsky}},\ }\href {\doibase 10.1103/PhysRevD.83.044026} {\bibfield
  {journal} {\bibinfo  {journal} {Phys. Rev. D}\ }\textbf {\bibinfo {volume}
  {83}},\ \bibinfo {pages} {044026} (\bibinfo {year} {2011})},\ \Eprint
  {http://arxiv.org/abs/1004.3558} {arXiv:1004.3558 [hep-th]} \BibitemShut
  {NoStop}%
\bibitem [{\citenamefont {Yoshino}\ and\ \citenamefont
  {Kodama}(2014)}]{Yoshino:2013ofa}%
  \BibitemOpen
  \bibfield  {author} {\bibinfo {author} {\bibfnamefont {H.}~\bibnamefont
  {Yoshino}}\ and\ \bibinfo {author} {\bibfnamefont {H.}~\bibnamefont
  {Kodama}},\ }\href {\doibase 10.1093/ptep/ptu029} {\bibfield  {journal}
  {\bibinfo  {journal} {PTEP}\ }\textbf {\bibinfo {volume} {2014}},\ \bibinfo
  {pages} {043E02} (\bibinfo {year} {2014})},\ \Eprint
  {http://arxiv.org/abs/1312.2326} {arXiv:1312.2326 [gr-qc]} \BibitemShut
  {NoStop}%
\bibitem [{\citenamefont {Chan}\ and\ \citenamefont
  {Hannuksela}(2024)}]{Chan:2022dkt}%
  \BibitemOpen
  \bibfield  {author} {\bibinfo {author} {\bibfnamefont {K.~H.~M.}\
  \bibnamefont {Chan}}\ and\ \bibinfo {author} {\bibfnamefont {O.~A.}\
  \bibnamefont {Hannuksela}},\ }\href {\doibase 10.1103/PhysRevD.109.023009}
  {\bibfield  {journal} {\bibinfo  {journal} {Phys. Rev. D}\ }\textbf {\bibinfo
  {volume} {109}},\ \bibinfo {pages} {023009} (\bibinfo {year} {2024})},\
  \Eprint {http://arxiv.org/abs/2209.03536} {arXiv:2209.03536 [gr-qc]}
  \BibitemShut {NoStop}%
\bibitem [{\citenamefont {Banerjee}\ \emph {et~al.}(2024)\citenamefont
  {Banerjee}, \citenamefont {Bonthu},\ and\ \citenamefont
  {Dey}}]{Banerjee:2024nga}%
  \BibitemOpen
  \bibfield  {author} {\bibinfo {author} {\bibfnamefont {I.~K.}\ \bibnamefont
  {Banerjee}}, \bibinfo {author} {\bibfnamefont {S.}~\bibnamefont {Bonthu}}, \
  and\ \bibinfo {author} {\bibfnamefont {U.~K.}\ \bibnamefont {Dey}},\
  }\href@noop {} {\  (\bibinfo {year} {2024})},\ \Eprint
  {http://arxiv.org/abs/2406.13756} {arXiv:2406.13756 [hep-ph]} \BibitemShut
  {NoStop}%
\bibitem [{\citenamefont {Brito}\ \emph
  {et~al.}(2017{\natexlab{a}})\citenamefont {Brito}, \citenamefont {Ghosh},
  \citenamefont {Barausse}, \citenamefont {Berti}, \citenamefont {Cardoso},
  \citenamefont {Dvorkin}, \citenamefont {Klein},\ and\ \citenamefont
  {Pani}}]{Brito:2017zvb}%
  \BibitemOpen
  \bibfield  {author} {\bibinfo {author} {\bibfnamefont {R.}~\bibnamefont
  {Brito}}, \bibinfo {author} {\bibfnamefont {S.}~\bibnamefont {Ghosh}},
  \bibinfo {author} {\bibfnamefont {E.}~\bibnamefont {Barausse}}, \bibinfo
  {author} {\bibfnamefont {E.}~\bibnamefont {Berti}}, \bibinfo {author}
  {\bibfnamefont {V.}~\bibnamefont {Cardoso}}, \bibinfo {author} {\bibfnamefont
  {I.}~\bibnamefont {Dvorkin}}, \bibinfo {author} {\bibfnamefont
  {A.}~\bibnamefont {Klein}}, \ and\ \bibinfo {author} {\bibfnamefont
  {P.}~\bibnamefont {Pani}},\ }\href {\doibase 10.1103/PhysRevD.96.064050}
  {\bibfield  {journal} {\bibinfo  {journal} {Phys. Rev. D}\ }\textbf {\bibinfo
  {volume} {96}},\ \bibinfo {pages} {064050} (\bibinfo {year}
  {2017}{\natexlab{a}})},\ \Eprint {http://arxiv.org/abs/1706.06311}
  {arXiv:1706.06311 [gr-qc]} \BibitemShut {NoStop}%
\bibitem [{\citenamefont {Brito}\ \emph
  {et~al.}(2017{\natexlab{b}})\citenamefont {Brito}, \citenamefont {Ghosh},
  \citenamefont {Barausse}, \citenamefont {Berti}, \citenamefont {Cardoso},
  \citenamefont {Dvorkin}, \citenamefont {Klein},\ and\ \citenamefont
  {Pani}}]{Brito:2017wnc}%
  \BibitemOpen
  \bibfield  {author} {\bibinfo {author} {\bibfnamefont {R.}~\bibnamefont
  {Brito}}, \bibinfo {author} {\bibfnamefont {S.}~\bibnamefont {Ghosh}},
  \bibinfo {author} {\bibfnamefont {E.}~\bibnamefont {Barausse}}, \bibinfo
  {author} {\bibfnamefont {E.}~\bibnamefont {Berti}}, \bibinfo {author}
  {\bibfnamefont {V.}~\bibnamefont {Cardoso}}, \bibinfo {author} {\bibfnamefont
  {I.}~\bibnamefont {Dvorkin}}, \bibinfo {author} {\bibfnamefont
  {A.}~\bibnamefont {Klein}}, \ and\ \bibinfo {author} {\bibfnamefont
  {P.}~\bibnamefont {Pani}},\ }\href {\doibase 10.1103/PhysRevLett.119.131101}
  {\bibfield  {journal} {\bibinfo  {journal} {Phys. Rev. Lett.}\ }\textbf
  {\bibinfo {volume} {119}},\ \bibinfo {pages} {131101} (\bibinfo {year}
  {2017}{\natexlab{b}})},\ \Eprint {http://arxiv.org/abs/1706.05097}
  {arXiv:1706.05097 [gr-qc]} \BibitemShut {NoStop}%
\bibitem [{\citenamefont {Isi}\ \emph {et~al.}(2019)\citenamefont {Isi},
  \citenamefont {Sun}, \citenamefont {Brito},\ and\ \citenamefont
  {Melatos}}]{Isi:2018pzk}%
  \BibitemOpen
  \bibfield  {author} {\bibinfo {author} {\bibfnamefont {M.}~\bibnamefont
  {Isi}}, \bibinfo {author} {\bibfnamefont {L.}~\bibnamefont {Sun}}, \bibinfo
  {author} {\bibfnamefont {R.}~\bibnamefont {Brito}}, \ and\ \bibinfo {author}
  {\bibfnamefont {A.}~\bibnamefont {Melatos}},\ }\href {\doibase
  10.1103/PhysRevD.99.084042} {\bibfield  {journal} {\bibinfo  {journal} {Phys.
  Rev. D}\ }\textbf {\bibinfo {volume} {99}},\ \bibinfo {pages} {084042}
  (\bibinfo {year} {2019})},\ \bibinfo {note} {[Erratum: Phys.Rev.D 102, 049901
  (2020)]},\ \Eprint {http://arxiv.org/abs/1810.03812} {arXiv:1810.03812
  [gr-qc]} \BibitemShut {NoStop}%
\bibitem [{\citenamefont {Ghosh}\ \emph {et~al.}(2019)\citenamefont {Ghosh},
  \citenamefont {Berti}, \citenamefont {Brito},\ and\ \citenamefont
  {Richartz}}]{Ghosh:2018gaw}%
  \BibitemOpen
  \bibfield  {author} {\bibinfo {author} {\bibfnamefont {S.}~\bibnamefont
  {Ghosh}}, \bibinfo {author} {\bibfnamefont {E.}~\bibnamefont {Berti}},
  \bibinfo {author} {\bibfnamefont {R.}~\bibnamefont {Brito}}, \ and\ \bibinfo
  {author} {\bibfnamefont {M.}~\bibnamefont {Richartz}},\ }\href {\doibase
  10.1103/PhysRevD.99.104030} {\bibfield  {journal} {\bibinfo  {journal} {Phys.
  Rev. D}\ }\textbf {\bibinfo {volume} {99}},\ \bibinfo {pages} {104030}
  (\bibinfo {year} {2019})},\ \Eprint {http://arxiv.org/abs/1812.01620}
  {arXiv:1812.01620 [gr-qc]} \BibitemShut {NoStop}%
\bibitem [{\citenamefont {Tsukada}\ \emph {et~al.}(2019)\citenamefont
  {Tsukada}, \citenamefont {Callister}, \citenamefont {Matas},\ and\
  \citenamefont {Meyers}}]{Tsukada:2018mbp}%
  \BibitemOpen
  \bibfield  {author} {\bibinfo {author} {\bibfnamefont {L.}~\bibnamefont
  {Tsukada}}, \bibinfo {author} {\bibfnamefont {T.}~\bibnamefont {Callister}},
  \bibinfo {author} {\bibfnamefont {A.}~\bibnamefont {Matas}}, \ and\ \bibinfo
  {author} {\bibfnamefont {P.}~\bibnamefont {Meyers}},\ }\href {\doibase
  10.1103/PhysRevD.99.103015} {\bibfield  {journal} {\bibinfo  {journal} {Phys.
  Rev. D}\ }\textbf {\bibinfo {volume} {99}},\ \bibinfo {pages} {103015}
  (\bibinfo {year} {2019})},\ \Eprint {http://arxiv.org/abs/1812.09622}
  {arXiv:1812.09622 [astro-ph.HE]} \BibitemShut {NoStop}%
\bibitem [{\citenamefont {Afshordi}\ \emph {et~al.}(2023)\citenamefont
  {Afshordi} \emph {et~al.}}]{LISAConsortiumWaveformWorkingGroup:2023arg}%
  \BibitemOpen
  \bibfield  {author} {\bibinfo {author} {\bibfnamefont {N.}~\bibnamefont
  {Afshordi}} \emph {et~al.} (\bibinfo {collaboration} {LISA Consortium
  Waveform Working Group}),\ }\href@noop {} {\  (\bibinfo {year} {2023})},\
  \Eprint {http://arxiv.org/abs/2311.01300} {arXiv:2311.01300 [gr-qc]}
  \BibitemShut {NoStop}%
\bibitem [{\citenamefont {Arvanitaki}\ \emph {et~al.}(2017)\citenamefont
  {Arvanitaki}, \citenamefont {Baryakhtar}, \citenamefont {Dimopoulos},
  \citenamefont {Dubovsky},\ and\ \citenamefont
  {Lasenby}}]{Arvanitaki:2016qwi}%
  \BibitemOpen
  \bibfield  {author} {\bibinfo {author} {\bibfnamefont {A.}~\bibnamefont
  {Arvanitaki}}, \bibinfo {author} {\bibfnamefont {M.}~\bibnamefont
  {Baryakhtar}}, \bibinfo {author} {\bibfnamefont {S.}~\bibnamefont
  {Dimopoulos}}, \bibinfo {author} {\bibfnamefont {S.}~\bibnamefont
  {Dubovsky}}, \ and\ \bibinfo {author} {\bibfnamefont {R.}~\bibnamefont
  {Lasenby}},\ }\href {\doibase 10.1103/PhysRevD.95.043001} {\bibfield
  {journal} {\bibinfo  {journal} {Phys. Rev. D}\ }\textbf {\bibinfo {volume}
  {95}},\ \bibinfo {pages} {043001} (\bibinfo {year} {2017})},\ \Eprint
  {http://arxiv.org/abs/1604.03958} {arXiv:1604.03958 [hep-ph]} \BibitemShut
  {NoStop}%
\bibitem [{\citenamefont {Ng}\ \emph {et~al.}(2021)\citenamefont {Ng},
  \citenamefont {Vitale}, \citenamefont {Hannuksela},\ and\ \citenamefont
  {Li}}]{Ng:2020ruv}%
  \BibitemOpen
  \bibfield  {author} {\bibinfo {author} {\bibfnamefont {K.~K.~Y.}\
  \bibnamefont {Ng}}, \bibinfo {author} {\bibfnamefont {S.}~\bibnamefont
  {Vitale}}, \bibinfo {author} {\bibfnamefont {O.~A.}\ \bibnamefont
  {Hannuksela}}, \ and\ \bibinfo {author} {\bibfnamefont {T.~G.~F.}\
  \bibnamefont {Li}},\ }\href {\doibase 10.1103/PhysRevLett.126.151102}
  {\bibfield  {journal} {\bibinfo  {journal} {Phys. Rev. Lett.}\ }\textbf
  {\bibinfo {volume} {126}},\ \bibinfo {pages} {151102} (\bibinfo {year}
  {2021})},\ \Eprint {http://arxiv.org/abs/2011.06010} {arXiv:2011.06010
  [gr-qc]} \BibitemShut {NoStop}%
\bibitem [{\citenamefont {Hui}\ \emph {et~al.}(2023)\citenamefont {Hui},
  \citenamefont {Law}, \citenamefont {Santoni}, \citenamefont {Sun},
  \citenamefont {Tomaselli},\ and\ \citenamefont {Trincherini}}]{Hui:2022sri}%
  \BibitemOpen
  \bibfield  {author} {\bibinfo {author} {\bibfnamefont {L.}~\bibnamefont
  {Hui}}, \bibinfo {author} {\bibfnamefont {Y.~T.~A.}\ \bibnamefont {Law}},
  \bibinfo {author} {\bibfnamefont {L.}~\bibnamefont {Santoni}}, \bibinfo
  {author} {\bibfnamefont {G.}~\bibnamefont {Sun}}, \bibinfo {author}
  {\bibfnamefont {G.~M.}\ \bibnamefont {Tomaselli}}, \ and\ \bibinfo {author}
  {\bibfnamefont {E.}~\bibnamefont {Trincherini}},\ }\href {\doibase
  10.1103/PhysRevD.107.104018} {\bibfield  {journal} {\bibinfo  {journal}
  {Phys. Rev. D}\ }\textbf {\bibinfo {volume} {107}},\ \bibinfo {pages}
  {104018} (\bibinfo {year} {2023})},\ \Eprint
  {http://arxiv.org/abs/2208.06408} {arXiv:2208.06408 [gr-qc]} \BibitemShut
  {NoStop}%
\bibitem [{\citenamefont {Ayzenberg}\ \emph {et~al.}(2023)\citenamefont
  {Ayzenberg} \emph {et~al.}}]{Ayzenberg:2023hfw}%
  \BibitemOpen
  \bibfield  {author} {\bibinfo {author} {\bibfnamefont {D.}~\bibnamefont
  {Ayzenberg}} \emph {et~al.},\ }\href@noop {} {\  (\bibinfo {year} {2023})},\
  \Eprint {http://arxiv.org/abs/2312.02130} {arXiv:2312.02130 [astro-ph.HE]}
  \BibitemShut {NoStop}%
\bibitem [{\citenamefont {Fukuda}\ and\ \citenamefont
  {Nakayama}(2020)}]{Fukuda:2019ewf}%
  \BibitemOpen
  \bibfield  {author} {\bibinfo {author} {\bibfnamefont {H.}~\bibnamefont
  {Fukuda}}\ and\ \bibinfo {author} {\bibfnamefont {K.}~\bibnamefont
  {Nakayama}},\ }\href {\doibase 10.1007/JHEP01(2020)128} {\bibfield  {journal}
  {\bibinfo  {journal} {JHEP}\ }\textbf {\bibinfo {volume} {01}},\ \bibinfo
  {pages} {128} (\bibinfo {year} {2020})},\ \Eprint
  {http://arxiv.org/abs/1910.06308} {arXiv:1910.06308 [hep-ph]} \BibitemShut
  {NoStop}%
\bibitem [{\citenamefont {Baryakhtar}\ \emph {et~al.}(2021)\citenamefont
  {Baryakhtar}, \citenamefont {Galanis}, \citenamefont {Lasenby},\ and\
  \citenamefont {Simon}}]{Baryakhtar:2020gao}%
  \BibitemOpen
  \bibfield  {author} {\bibinfo {author} {\bibfnamefont {M.}~\bibnamefont
  {Baryakhtar}}, \bibinfo {author} {\bibfnamefont {M.}~\bibnamefont {Galanis}},
  \bibinfo {author} {\bibfnamefont {R.}~\bibnamefont {Lasenby}}, \ and\
  \bibinfo {author} {\bibfnamefont {O.}~\bibnamefont {Simon}},\ }\href
  {\doibase 10.1103/PhysRevD.103.095019} {\bibfield  {journal} {\bibinfo
  {journal} {Phys. Rev. D}\ }\textbf {\bibinfo {volume} {103}},\ \bibinfo
  {pages} {095019} (\bibinfo {year} {2021})},\ \Eprint
  {http://arxiv.org/abs/2011.11646} {arXiv:2011.11646 [hep-ph]} \BibitemShut
  {NoStop}%
\bibitem [{\citenamefont {Omiya}\ \emph {et~al.}(2022)\citenamefont {Omiya},
  \citenamefont {Takahashi},\ and\ \citenamefont {Tanaka}}]{Omiya:2022mwv}%
  \BibitemOpen
  \bibfield  {author} {\bibinfo {author} {\bibfnamefont {H.}~\bibnamefont
  {Omiya}}, \bibinfo {author} {\bibfnamefont {T.}~\bibnamefont {Takahashi}}, \
  and\ \bibinfo {author} {\bibfnamefont {T.}~\bibnamefont {Tanaka}},\ }\href
  {\doibase 10.1093/ptep/ptac058} {\bibfield  {journal} {\bibinfo  {journal}
  {PTEP}\ }\textbf {\bibinfo {volume} {2022}},\ \bibinfo {pages} {043E03}
  (\bibinfo {year} {2022})},\ \Eprint {http://arxiv.org/abs/2201.04382}
  {arXiv:2201.04382 [gr-qc]} \BibitemShut {NoStop}%
\bibitem [{\citenamefont {Takahashi}\ \emph {et~al.}(2024)\citenamefont
  {Takahashi}, \citenamefont {Omiya},\ and\ \citenamefont
  {Tanaka}}]{Takahashi:2024fyq}%
  \BibitemOpen
  \bibfield  {author} {\bibinfo {author} {\bibfnamefont {T.}~\bibnamefont
  {Takahashi}}, \bibinfo {author} {\bibfnamefont {H.}~\bibnamefont {Omiya}}, \
  and\ \bibinfo {author} {\bibfnamefont {T.}~\bibnamefont {Tanaka}},\
  }\href@noop {} {\  (\bibinfo {year} {2024})},\ \Eprint
  {http://arxiv.org/abs/2408.08349} {arXiv:2408.08349 [gr-qc]} \BibitemShut
  {NoStop}%
\bibitem [{\citenamefont {Collaviti}\ \emph {et~al.}(2024)\citenamefont
  {Collaviti}, \citenamefont {Sun}, \citenamefont {Galanis},\ and\
  \citenamefont {Baryakhtar}}]{Collaviti:2024mvh}%
  \BibitemOpen
  \bibfield  {author} {\bibinfo {author} {\bibfnamefont {S.}~\bibnamefont
  {Collaviti}}, \bibinfo {author} {\bibfnamefont {L.}~\bibnamefont {Sun}},
  \bibinfo {author} {\bibfnamefont {M.}~\bibnamefont {Galanis}}, \ and\
  \bibinfo {author} {\bibfnamefont {M.}~\bibnamefont {Baryakhtar}},\
  }\href@noop {} {\  (\bibinfo {year} {2024})},\ \Eprint
  {http://arxiv.org/abs/2407.04304} {arXiv:2407.04304 [gr-qc]} \BibitemShut
  {NoStop}%
\bibitem [{\citenamefont {Hoof}\ \emph {et~al.}(2024)\citenamefont {Hoof},
  \citenamefont {Marsh}, \citenamefont {Sisk-Reyn\'es}, \citenamefont
  {Matthews},\ and\ \citenamefont {Reynolds}}]{Hoof:2024quk}%
  \BibitemOpen
  \bibfield  {author} {\bibinfo {author} {\bibfnamefont {S.}~\bibnamefont
  {Hoof}}, \bibinfo {author} {\bibfnamefont {D.~J.~E.}\ \bibnamefont {Marsh}},
  \bibinfo {author} {\bibfnamefont {J.}~\bibnamefont {Sisk-Reyn\'es}}, \bibinfo
  {author} {\bibfnamefont {J.~H.}\ \bibnamefont {Matthews}}, \ and\ \bibinfo
  {author} {\bibfnamefont {C.}~\bibnamefont {Reynolds}},\ }\href@noop {} {\
  (\bibinfo {year} {2024})},\ \Eprint {http://arxiv.org/abs/2406.10337}
  {arXiv:2406.10337 [hep-ph]} \BibitemShut {NoStop}%
\bibitem [{\citenamefont {Baumann}\ \emph {et~al.}(2020)\citenamefont
  {Baumann}, \citenamefont {Chia}, \citenamefont {Porto},\ and\ \citenamefont
  {Stout}}]{Baumann:2019ztm}%
  \BibitemOpen
  \bibfield  {author} {\bibinfo {author} {\bibfnamefont {D.}~\bibnamefont
  {Baumann}}, \bibinfo {author} {\bibfnamefont {H.~S.}\ \bibnamefont {Chia}},
  \bibinfo {author} {\bibfnamefont {R.~A.}\ \bibnamefont {Porto}}, \ and\
  \bibinfo {author} {\bibfnamefont {J.}~\bibnamefont {Stout}},\ }\href
  {\doibase 10.1103/PhysRevD.101.083019} {\bibfield  {journal} {\bibinfo
  {journal} {Phys. Rev. D}\ }\textbf {\bibinfo {volume} {101}},\ \bibinfo
  {pages} {083019} (\bibinfo {year} {2020})},\ \Eprint
  {http://arxiv.org/abs/1912.04932} {arXiv:1912.04932 [gr-qc]} \BibitemShut
  {NoStop}%
\bibitem [{\citenamefont {Baumann}\ \emph
  {et~al.}(2019{\natexlab{a}})\citenamefont {Baumann}, \citenamefont {Chia},\
  and\ \citenamefont {Porto}}]{Baumann:2018vus}%
  \BibitemOpen
  \bibfield  {author} {\bibinfo {author} {\bibfnamefont {D.}~\bibnamefont
  {Baumann}}, \bibinfo {author} {\bibfnamefont {H.~S.}\ \bibnamefont {Chia}}, \
  and\ \bibinfo {author} {\bibfnamefont {R.~A.}\ \bibnamefont {Porto}},\ }\href
  {\doibase 10.1103/PhysRevD.99.044001} {\bibfield  {journal} {\bibinfo
  {journal} {Phys. Rev. D}\ }\textbf {\bibinfo {volume} {99}},\ \bibinfo
  {pages} {044001} (\bibinfo {year} {2019}{\natexlab{a}})},\ \Eprint
  {http://arxiv.org/abs/1804.03208} {arXiv:1804.03208 [gr-qc]} \BibitemShut
  {NoStop}%
\bibitem [{\citenamefont {Baumann}\ \emph
  {et~al.}(2019{\natexlab{b}})\citenamefont {Baumann}, \citenamefont {Chia},
  \citenamefont {Stout},\ and\ \citenamefont {ter Haar}}]{Baumann:2019eav}%
  \BibitemOpen
  \bibfield  {author} {\bibinfo {author} {\bibfnamefont {D.}~\bibnamefont
  {Baumann}}, \bibinfo {author} {\bibfnamefont {H.~S.}\ \bibnamefont {Chia}},
  \bibinfo {author} {\bibfnamefont {J.}~\bibnamefont {Stout}}, \ and\ \bibinfo
  {author} {\bibfnamefont {L.}~\bibnamefont {ter Haar}},\ }\href {\doibase
  10.1088/1475-7516/2019/12/006} {\bibfield  {journal} {\bibinfo  {journal}
  {JCAP}\ }\textbf {\bibinfo {volume} {12}},\ \bibinfo {pages} {006} (\bibinfo
  {year} {2019}{\natexlab{b}})},\ \Eprint {http://arxiv.org/abs/1908.10370}
  {arXiv:1908.10370 [gr-qc]} \BibitemShut {NoStop}%
\bibitem [{\citenamefont {Ding}\ \emph {et~al.}(2021)\citenamefont {Ding},
  \citenamefont {Tong},\ and\ \citenamefont {Wang}}]{Ding:2020bnl}%
  \BibitemOpen
  \bibfield  {author} {\bibinfo {author} {\bibfnamefont {Q.}~\bibnamefont
  {Ding}}, \bibinfo {author} {\bibfnamefont {X.}~\bibnamefont {Tong}}, \ and\
  \bibinfo {author} {\bibfnamefont {Y.}~\bibnamefont {Wang}},\ }\href {\doibase
  10.3847/1538-4357/abd803} {\bibfield  {journal} {\bibinfo  {journal}
  {Astrophys. J.}\ }\textbf {\bibinfo {volume} {908}},\ \bibinfo {pages} {78}
  (\bibinfo {year} {2021})},\ \Eprint {http://arxiv.org/abs/2009.11106}
  {arXiv:2009.11106 [astro-ph.HE]} \BibitemShut {NoStop}%
\bibitem [{\citenamefont {Tong}\ \emph
  {et~al.}(2022{\natexlab{a}})\citenamefont {Tong}, \citenamefont {Wang},\ and\
  \citenamefont {Zhu}}]{Tong:2021whq}%
  \BibitemOpen
  \bibfield  {author} {\bibinfo {author} {\bibfnamefont {X.}~\bibnamefont
  {Tong}}, \bibinfo {author} {\bibfnamefont {Y.}~\bibnamefont {Wang}}, \ and\
  \bibinfo {author} {\bibfnamefont {H.-Y.}\ \bibnamefont {Zhu}},\ }\href
  {\doibase 10.3847/1538-4357/ac36db} {\bibfield  {journal} {\bibinfo
  {journal} {Astrophys. J.}\ }\textbf {\bibinfo {volume} {924}},\ \bibinfo
  {pages} {99} (\bibinfo {year} {2022}{\natexlab{a}})},\ \Eprint
  {http://arxiv.org/abs/2106.13484} {arXiv:2106.13484 [astro-ph.HE]}
  \BibitemShut {NoStop}%
\bibitem [{\citenamefont {Arvanitaki}\ \emph {et~al.}(2015)\citenamefont
  {Arvanitaki}, \citenamefont {Baryakhtar},\ and\ \citenamefont
  {Huang}}]{Arvanitaki:2014wva}%
  \BibitemOpen
  \bibfield  {author} {\bibinfo {author} {\bibfnamefont {A.}~\bibnamefont
  {Arvanitaki}}, \bibinfo {author} {\bibfnamefont {M.}~\bibnamefont
  {Baryakhtar}}, \ and\ \bibinfo {author} {\bibfnamefont {X.}~\bibnamefont
  {Huang}},\ }\href {\doibase 10.1103/PhysRevD.91.084011} {\bibfield  {journal}
  {\bibinfo  {journal} {Phys. Rev. D}\ }\textbf {\bibinfo {volume} {91}},\
  \bibinfo {pages} {084011} (\bibinfo {year} {2015})},\ \Eprint
  {http://arxiv.org/abs/1411.2263} {arXiv:1411.2263 [hep-ph]} \BibitemShut
  {NoStop}%
\bibitem [{\citenamefont {Du}\ \emph {et~al.}(2022)\citenamefont {Du},
  \citenamefont {Egana-Ugrinovic}, \citenamefont {Essig}, \citenamefont
  {Fragione},\ and\ \citenamefont {Perna}}]{Du:2022trq}%
  \BibitemOpen
  \bibfield  {author} {\bibinfo {author} {\bibfnamefont {P.}~\bibnamefont
  {Du}}, \bibinfo {author} {\bibfnamefont {D.}~\bibnamefont {Egana-Ugrinovic}},
  \bibinfo {author} {\bibfnamefont {R.}~\bibnamefont {Essig}}, \bibinfo
  {author} {\bibfnamefont {G.}~\bibnamefont {Fragione}}, \ and\ \bibinfo
  {author} {\bibfnamefont {R.}~\bibnamefont {Perna}},\ }\href {\doibase
  10.1038/s41467-022-32301-4} {\bibfield  {journal} {\bibinfo  {journal}
  {Nature Commun.}\ }\textbf {\bibinfo {volume} {13}},\ \bibinfo {pages} {4626}
  (\bibinfo {year} {2022})},\ \Eprint {http://arxiv.org/abs/2202.01215}
  {arXiv:2202.01215 [hep-ph]} \BibitemShut {NoStop}%
\bibitem [{\citenamefont {Tong}\ \emph
  {et~al.}(2022{\natexlab{b}})\citenamefont {Tong}, \citenamefont {Wang},\ and\
  \citenamefont {Zhu}}]{Tong:2022bbl}%
  \BibitemOpen
  \bibfield  {author} {\bibinfo {author} {\bibfnamefont {X.}~\bibnamefont
  {Tong}}, \bibinfo {author} {\bibfnamefont {Y.}~\bibnamefont {Wang}}, \ and\
  \bibinfo {author} {\bibfnamefont {H.-Y.}\ \bibnamefont {Zhu}},\ }\href
  {\doibase 10.1103/PhysRevD.106.043002} {\enquote {\bibinfo {title}
  {{Termination of superradiance from a binary companion}},}\ } (\bibinfo
  {year} {2022}{\natexlab{b}}),\ \Eprint {http://arxiv.org/abs/2205.10527}
  {arXiv:2205.10527 [gr-qc]} \BibitemShut {NoStop}%
\bibitem [{\citenamefont {Fan}\ \emph {et~al.}(2024)\citenamefont {Fan},
  \citenamefont {Tong}, \citenamefont {Wang},\ and\ \citenamefont
  {Zhu}}]{Fan:2023jjj}%
  \BibitemOpen
  \bibfield  {author} {\bibinfo {author} {\bibfnamefont {K.}~\bibnamefont
  {Fan}}, \bibinfo {author} {\bibfnamefont {X.}~\bibnamefont {Tong}}, \bibinfo
  {author} {\bibfnamefont {Y.}~\bibnamefont {Wang}}, \ and\ \bibinfo {author}
  {\bibfnamefont {H.-Y.}\ \bibnamefont {Zhu}},\ }\href {\doibase
  10.1103/PhysRevD.109.024059} {\bibfield  {journal} {\bibinfo  {journal}
  {Phys. Rev. D}\ }\textbf {\bibinfo {volume} {109}},\ \bibinfo {pages}
  {024059} (\bibinfo {year} {2024})},\ \Eprint
  {http://arxiv.org/abs/2311.17013} {arXiv:2311.17013 [gr-qc]} \BibitemShut
  {NoStop}%
\bibitem [{\citenamefont {{Lam}}\ \emph {et~al.}(2022)\citenamefont {{Lam}},
  \citenamefont {{Lu}}, \citenamefont {{Udalski}}, \citenamefont {{Bond}},
  \citenamefont {{Bennett}}, \citenamefont {{Skowron}}, \citenamefont
  {{Mr{\'o}z}}, \citenamefont {{Poleski}}, \citenamefont {{Sumi}},
  \citenamefont {{Szyma{\'n}ski}}, \citenamefont {{Koz{\l}owski}},
  \citenamefont {{Pietrukowicz}}, \citenamefont {{Soszy{\'n}ski}},
  \citenamefont {{Ulaczyk}}, \citenamefont {{Wyrzykowski}}, \citenamefont
  {{Miyazaki}}, \citenamefont {{Suzuki}}, \citenamefont {{Koshimoto}},
  \citenamefont {{Rattenbury}}, \citenamefont {{Hosek}}, \citenamefont {{Abe}},
  \citenamefont {{Barry}}, \citenamefont {{Bhattacharya}}, \citenamefont
  {{Fukui}}, \citenamefont {{Fujii}}, \citenamefont {{Hirao}}, \citenamefont
  {{Itow}}, \citenamefont {{Kirikawa}}, \citenamefont {{Kondo}}, \citenamefont
  {{Matsubara}}, \citenamefont {{Matsumoto}}, \citenamefont {{Muraki}},
  \citenamefont {{Olmschenk}}, \citenamefont {{Ranc}}, \citenamefont
  {{Okamura}}, \citenamefont {{Satoh}}, \citenamefont {{Silva}}, \citenamefont
  {{Toda}}, \citenamefont {{Tristram}}, \citenamefont {{Vandorou}},
  \citenamefont {{Yama}}, \citenamefont {{Abrams}}, \citenamefont {{Agarwal}},
  \citenamefont {{Rose}},\ and\ \citenamefont {{Terry}}}]{Lam:2022}%
  \BibitemOpen
  \bibfield  {author} {\bibinfo {author} {\bibfnamefont {C.~Y.}\ \bibnamefont
  {{Lam}}}, \bibinfo {author} {\bibfnamefont {J.~R.}\ \bibnamefont {{Lu}}},
  \bibinfo {author} {\bibfnamefont {A.}~\bibnamefont {{Udalski}}}, \bibinfo
  {author} {\bibfnamefont {I.}~\bibnamefont {{Bond}}}, \bibinfo {author}
  {\bibfnamefont {D.~P.}\ \bibnamefont {{Bennett}}}, \bibinfo {author}
  {\bibfnamefont {J.}~\bibnamefont {{Skowron}}}, \bibinfo {author}
  {\bibfnamefont {P.}~\bibnamefont {{Mr{\'o}z}}}, \bibinfo {author}
  {\bibfnamefont {R.}~\bibnamefont {{Poleski}}}, \bibinfo {author}
  {\bibfnamefont {T.}~\bibnamefont {{Sumi}}}, \bibinfo {author} {\bibfnamefont
  {M.~K.}\ \bibnamefont {{Szyma{\'n}ski}}}, \bibinfo {author} {\bibfnamefont
  {S.}~\bibnamefont {{Koz{\l}owski}}}, \bibinfo {author} {\bibfnamefont
  {P.}~\bibnamefont {{Pietrukowicz}}}, \bibinfo {author} {\bibfnamefont
  {I.}~\bibnamefont {{Soszy{\'n}ski}}}, \bibinfo {author} {\bibfnamefont
  {K.}~\bibnamefont {{Ulaczyk}}}, \bibinfo {author} {\bibfnamefont
  {{\L}.}~\bibnamefont {{Wyrzykowski}}}, \bibinfo {author} {\bibfnamefont
  {S.}~\bibnamefont {{Miyazaki}}}, \bibinfo {author} {\bibfnamefont
  {D.}~\bibnamefont {{Suzuki}}}, \bibinfo {author} {\bibfnamefont
  {N.}~\bibnamefont {{Koshimoto}}}, \bibinfo {author} {\bibfnamefont {N.~J.}\
  \bibnamefont {{Rattenbury}}}, \bibinfo {author} {\bibfnamefont {M.~W.}\
  \bibnamefont {{Hosek}}}, \bibinfo {author} {\bibfnamefont {F.}~\bibnamefont
  {{Abe}}}, \bibinfo {author} {\bibfnamefont {R.}~\bibnamefont {{Barry}}},
  \bibinfo {author} {\bibfnamefont {A.}~\bibnamefont {{Bhattacharya}}},
  \bibinfo {author} {\bibfnamefont {A.}~\bibnamefont {{Fukui}}}, \bibinfo
  {author} {\bibfnamefont {H.}~\bibnamefont {{Fujii}}}, \bibinfo {author}
  {\bibfnamefont {Y.}~\bibnamefont {{Hirao}}}, \bibinfo {author} {\bibfnamefont
  {Y.}~\bibnamefont {{Itow}}}, \bibinfo {author} {\bibfnamefont
  {R.}~\bibnamefont {{Kirikawa}}}, \bibinfo {author} {\bibfnamefont
  {I.}~\bibnamefont {{Kondo}}}, \bibinfo {author} {\bibfnamefont
  {Y.}~\bibnamefont {{Matsubara}}}, \bibinfo {author} {\bibfnamefont
  {S.}~\bibnamefont {{Matsumoto}}}, \bibinfo {author} {\bibfnamefont
  {Y.}~\bibnamefont {{Muraki}}}, \bibinfo {author} {\bibfnamefont
  {G.}~\bibnamefont {{Olmschenk}}}, \bibinfo {author} {\bibfnamefont
  {C.}~\bibnamefont {{Ranc}}}, \bibinfo {author} {\bibfnamefont
  {A.}~\bibnamefont {{Okamura}}}, \bibinfo {author} {\bibfnamefont
  {Y.}~\bibnamefont {{Satoh}}}, \bibinfo {author} {\bibfnamefont {S.~I.}\
  \bibnamefont {{Silva}}}, \bibinfo {author} {\bibfnamefont {T.}~\bibnamefont
  {{Toda}}}, \bibinfo {author} {\bibfnamefont {P.~J.}\ \bibnamefont
  {{Tristram}}}, \bibinfo {author} {\bibfnamefont {A.}~\bibnamefont
  {{Vandorou}}}, \bibinfo {author} {\bibfnamefont {H.}~\bibnamefont {{Yama}}},
  \bibinfo {author} {\bibfnamefont {N.~S.}\ \bibnamefont {{Abrams}}}, \bibinfo
  {author} {\bibfnamefont {S.}~\bibnamefont {{Agarwal}}}, \bibinfo {author}
  {\bibfnamefont {S.}~\bibnamefont {{Rose}}}, \ and\ \bibinfo {author}
  {\bibfnamefont {S.~K.}\ \bibnamefont {{Terry}}},\ }\href {\doibase
  10.3847/2041-8213/ac7442} {\bibfield  {journal} {\bibinfo  {journal} {\apjl}\
  }\textbf {\bibinfo {volume} {933}},\ \bibinfo {eid} {L23} (\bibinfo {year}
  {2022})},\ \Eprint {http://arxiv.org/abs/2202.01903} {arXiv:2202.01903
  [astro-ph.GA]} \BibitemShut {NoStop}%
\bibitem [{\citenamefont {{Belczynski}}\ \emph {et~al.}(2004)\citenamefont
  {{Belczynski}}, \citenamefont {{Sadowski}},\ and\ \citenamefont
  {{Rasio}}}]{Belczynski:2004}%
  \BibitemOpen
  \bibfield  {author} {\bibinfo {author} {\bibfnamefont {K.}~\bibnamefont
  {{Belczynski}}}, \bibinfo {author} {\bibfnamefont {A.}~\bibnamefont
  {{Sadowski}}}, \ and\ \bibinfo {author} {\bibfnamefont {F.~A.}\ \bibnamefont
  {{Rasio}}},\ }\href {\doibase 10.1086/422191} {\bibfield  {journal} {\bibinfo
   {journal} {\apj}\ }\textbf {\bibinfo {volume} {611}},\ \bibinfo {pages}
  {1068} (\bibinfo {year} {2004})},\ \Eprint
  {http://arxiv.org/abs/astro-ph/0404068} {arXiv:astro-ph/0404068 [astro-ph]}
  \BibitemShut {NoStop}%
\bibitem [{\citenamefont {{Fender}}\ \emph {et~al.}(2013)\citenamefont
  {{Fender}}, \citenamefont {{Maccarone}},\ and\ \citenamefont
  {{Heywood}}}]{Fender:2013}%
  \BibitemOpen
  \bibfield  {author} {\bibinfo {author} {\bibfnamefont {R.~P.}\ \bibnamefont
  {{Fender}}}, \bibinfo {author} {\bibfnamefont {T.~J.}\ \bibnamefont
  {{Maccarone}}}, \ and\ \bibinfo {author} {\bibfnamefont {I.}~\bibnamefont
  {{Heywood}}},\ }\href {\doibase 10.1093/mnras/sts688} {\bibfield  {journal}
  {\bibinfo  {journal} {\mnras}\ }\textbf {\bibinfo {volume} {430}},\ \bibinfo
  {pages} {1538} (\bibinfo {year} {2013})},\ \Eprint
  {http://arxiv.org/abs/1301.1341} {arXiv:1301.1341 [astro-ph.HE]} \BibitemShut
  {NoStop}%
\bibitem [{\citenamefont {{Wiktorowicz}}\ \emph {et~al.}(2019)\citenamefont
  {{Wiktorowicz}}, \citenamefont {{Wyrzykowski}}, \citenamefont
  {{Chruslinska}}, \citenamefont {{Klencki}}, \citenamefont {{Rybicki}},\ and\
  \citenamefont {{Belczynski}}}]{Wiktorowicz:2019}%
  \BibitemOpen
  \bibfield  {author} {\bibinfo {author} {\bibfnamefont {G.}~\bibnamefont
  {{Wiktorowicz}}}, \bibinfo {author} {\bibfnamefont {{\L}.}~\bibnamefont
  {{Wyrzykowski}}}, \bibinfo {author} {\bibfnamefont {M.}~\bibnamefont
  {{Chruslinska}}}, \bibinfo {author} {\bibfnamefont {J.}~\bibnamefont
  {{Klencki}}}, \bibinfo {author} {\bibfnamefont {K.~A.}\ \bibnamefont
  {{Rybicki}}}, \ and\ \bibinfo {author} {\bibfnamefont {K.}~\bibnamefont
  {{Belczynski}}},\ }\href {\doibase 10.3847/1538-4357/ab45e6} {\bibfield
  {journal} {\bibinfo  {journal} {\apj}\ }\textbf {\bibinfo {volume} {885}},\
  \bibinfo {eid} {1} (\bibinfo {year} {2019})},\ \Eprint
  {http://arxiv.org/abs/1907.11431} {arXiv:1907.11431 [astro-ph.HE]}
  \BibitemShut {NoStop}%
\bibitem [{\citenamefont {{Gou}}\ \emph {et~al.}(2011)\citenamefont {{Gou}},
  \citenamefont {{McClintock}}, \citenamefont {{Reid}}, \citenamefont
  {{Orosz}}, \citenamefont {{Steiner}}, \citenamefont {{Narayan}},
  \citenamefont {{Xiang}}, \citenamefont {{Remillard}}, \citenamefont
  {{Arnaud}},\ and\ \citenamefont {{Davis}}}]{Gou:2011}%
  \BibitemOpen
  \bibfield  {author} {\bibinfo {author} {\bibfnamefont {L.}~\bibnamefont
  {{Gou}}}, \bibinfo {author} {\bibfnamefont {J.~E.}\ \bibnamefont
  {{McClintock}}}, \bibinfo {author} {\bibfnamefont {M.~J.}\ \bibnamefont
  {{Reid}}}, \bibinfo {author} {\bibfnamefont {J.~A.}\ \bibnamefont {{Orosz}}},
  \bibinfo {author} {\bibfnamefont {J.~F.}\ \bibnamefont {{Steiner}}}, \bibinfo
  {author} {\bibfnamefont {R.}~\bibnamefont {{Narayan}}}, \bibinfo {author}
  {\bibfnamefont {J.}~\bibnamefont {{Xiang}}}, \bibinfo {author} {\bibfnamefont
  {R.~A.}\ \bibnamefont {{Remillard}}}, \bibinfo {author} {\bibfnamefont
  {K.~A.}\ \bibnamefont {{Arnaud}}}, \ and\ \bibinfo {author} {\bibfnamefont
  {S.~W.}\ \bibnamefont {{Davis}}},\ }\href {\doibase
  10.1088/0004-637X/742/2/85} {\bibfield  {journal} {\bibinfo  {journal}
  {\apj}\ }\textbf {\bibinfo {volume} {742}},\ \bibinfo {eid} {85} (\bibinfo
  {year} {2011})},\ \Eprint {http://arxiv.org/abs/1106.3690} {arXiv:1106.3690
  [astro-ph.HE]} \BibitemShut {NoStop}%
\bibitem [{\citenamefont {{McClintock}}\ \emph {et~al.}(2014)\citenamefont
  {{McClintock}}, \citenamefont {{Narayan}},\ and\ \citenamefont
  {{Steiner}}}]{McClintock:2014}%
  \BibitemOpen
  \bibfield  {author} {\bibinfo {author} {\bibfnamefont {J.~E.}\ \bibnamefont
  {{McClintock}}}, \bibinfo {author} {\bibfnamefont {R.}~\bibnamefont
  {{Narayan}}}, \ and\ \bibinfo {author} {\bibfnamefont {J.~F.}\ \bibnamefont
  {{Steiner}}},\ }\href {\doibase 10.1007/s11214-013-0003-9} {\bibfield
  {journal} {\bibinfo  {journal} {\ssr}\ }\textbf {\bibinfo {volume} {183}},\
  \bibinfo {pages} {295} (\bibinfo {year} {2014})},\ \Eprint
  {http://arxiv.org/abs/1303.1583} {arXiv:1303.1583 [astro-ph.HE]} \BibitemShut
  {NoStop}%
\bibitem [{\citenamefont {Fuller}\ and\ \citenamefont
  {Ma}(2019)}]{Fuller:2019sxi}%
  \BibitemOpen
  \bibfield  {author} {\bibinfo {author} {\bibfnamefont {J.}~\bibnamefont
  {Fuller}}\ and\ \bibinfo {author} {\bibfnamefont {L.}~\bibnamefont {Ma}},\
  }\href {\doibase 10.3847/2041-8213/ab339b} {\bibfield  {journal} {\bibinfo
  {journal} {Astrophys. J. Lett.}\ }\textbf {\bibinfo {volume} {881}},\
  \bibinfo {pages} {L1} (\bibinfo {year} {2019})},\ \Eprint
  {http://arxiv.org/abs/1907.03714} {arXiv:1907.03714 [astro-ph.SR]}
  \BibitemShut {NoStop}%
\bibitem [{\citenamefont {Siemonsen}\ \emph {et~al.}(2023)\citenamefont
  {Siemonsen}, \citenamefont {May},\ and\ \citenamefont
  {East}}]{Siemonsen:2022yyf}%
  \BibitemOpen
  \bibfield  {author} {\bibinfo {author} {\bibfnamefont {N.}~\bibnamefont
  {Siemonsen}}, \bibinfo {author} {\bibfnamefont {T.}~\bibnamefont {May}}, \
  and\ \bibinfo {author} {\bibfnamefont {W.~E.}\ \bibnamefont {East}},\ }\href
  {\doibase 10.1103/PhysRevD.107.104003} {\bibfield  {journal} {\bibinfo
  {journal} {Phys. Rev. D}\ }\textbf {\bibinfo {volume} {107}},\ \bibinfo
  {pages} {104003} (\bibinfo {year} {2023})},\ \Eprint
  {http://arxiv.org/abs/2211.03845} {arXiv:2211.03845 [gr-qc]} \BibitemShut
  {NoStop}%
\bibitem [{\citenamefont {Brito}\ \emph
  {et~al.}(2015{\natexlab{b}})\citenamefont {Brito}, \citenamefont {Cardoso},\
  and\ \citenamefont {Pani}}]{Brito:2014wla}%
  \BibitemOpen
  \bibfield  {author} {\bibinfo {author} {\bibfnamefont {R.}~\bibnamefont
  {Brito}}, \bibinfo {author} {\bibfnamefont {V.}~\bibnamefont {Cardoso}}, \
  and\ \bibinfo {author} {\bibfnamefont {P.}~\bibnamefont {Pani}},\ }\href
  {\doibase 10.1088/0264-9381/32/13/134001} {\bibfield  {journal} {\bibinfo
  {journal} {Class. Quant. Grav.}\ }\textbf {\bibinfo {volume} {32}},\ \bibinfo
  {pages} {134001} (\bibinfo {year} {2015}{\natexlab{b}})},\ \Eprint
  {http://arxiv.org/abs/1411.0686} {arXiv:1411.0686 [gr-qc]} \BibitemShut
  {NoStop}%
\bibitem [{\citenamefont {Spera}\ \emph {et~al.}(2015)\citenamefont {Spera},
  \citenamefont {Mapelli},\ and\ \citenamefont {Bressan}}]{spera2015mass}%
  \BibitemOpen
  \bibfield  {author} {\bibinfo {author} {\bibfnamefont {M.}~\bibnamefont
  {Spera}}, \bibinfo {author} {\bibfnamefont {M.}~\bibnamefont {Mapelli}}, \
  and\ \bibinfo {author} {\bibfnamefont {A.}~\bibnamefont {Bressan}},\
  }\href@noop {} {\bibfield  {journal} {\bibinfo  {journal} {Monthly Notices of
  the Royal Astronomical Society}\ }\textbf {\bibinfo {volume} {451}},\
  \bibinfo {pages} {4086} (\bibinfo {year} {2015})}\BibitemShut {NoStop}%
\bibitem [{\citenamefont {Spera}\ and\ \citenamefont
  {Mapelli}(2017)}]{Spera:2017fyx}%
  \BibitemOpen
  \bibfield  {author} {\bibinfo {author} {\bibfnamefont {M.}~\bibnamefont
  {Spera}}\ and\ \bibinfo {author} {\bibfnamefont {M.}~\bibnamefont
  {Mapelli}},\ }\href {\doibase 10.1093/mnras/stx1576} {\bibfield  {journal}
  {\bibinfo  {journal} {Mon. Not. Roy. Astron. Soc.}\ }\textbf {\bibinfo
  {volume} {470}},\ \bibinfo {pages} {4739} (\bibinfo {year} {2017})},\ \Eprint
  {http://arxiv.org/abs/1706.06109} {arXiv:1706.06109 [astro-ph.SR]}
  \BibitemShut {NoStop}%
\bibitem [{\citenamefont {Spera}\ \emph {et~al.}(2018)\citenamefont {Spera},
  \citenamefont {Mapelli}, \citenamefont {Giacobbo}, \citenamefont {Trani},
  \citenamefont {Bressan},\ and\ \citenamefont {Costa}}]{Spera:2018wnw}%
  \BibitemOpen
  \bibfield  {author} {\bibinfo {author} {\bibfnamefont {M.}~\bibnamefont
  {Spera}}, \bibinfo {author} {\bibfnamefont {M.}~\bibnamefont {Mapelli}},
  \bibinfo {author} {\bibfnamefont {N.}~\bibnamefont {Giacobbo}}, \bibinfo
  {author} {\bibfnamefont {A.~A.}\ \bibnamefont {Trani}}, \bibinfo {author}
  {\bibfnamefont {A.}~\bibnamefont {Bressan}}, \ and\ \bibinfo {author}
  {\bibfnamefont {G.}~\bibnamefont {Costa}},\ }\href {\doibase
  10.1093/mnras/stz359} {\  (\bibinfo {year} {2018}),\ 10.1093/mnras/stz359},\
  \Eprint {http://arxiv.org/abs/1809.04605} {arXiv:1809.04605 [astro-ph.HE]}
  \BibitemShut {NoStop}%
\bibitem [{\citenamefont {Sana}\ \emph {et~al.}(2012)\citenamefont {Sana},
  \citenamefont {de~Mink}, \citenamefont {de~Koter}, \citenamefont {Langer},
  \citenamefont {Evans}, \citenamefont {Gieles}, \citenamefont {Gosset},
  \citenamefont {Izzard}, \citenamefont {Bouquin},\ and\ \citenamefont
  {Schneider}}]{Sana:2012px}%
  \BibitemOpen
  \bibfield  {author} {\bibinfo {author} {\bibfnamefont {H.}~\bibnamefont
  {Sana}}, \bibinfo {author} {\bibfnamefont {S.~E.}\ \bibnamefont {de~Mink}},
  \bibinfo {author} {\bibfnamefont {A.}~\bibnamefont {de~Koter}}, \bibinfo
  {author} {\bibfnamefont {N.}~\bibnamefont {Langer}}, \bibinfo {author}
  {\bibfnamefont {C.~J.}\ \bibnamefont {Evans}}, \bibinfo {author}
  {\bibfnamefont {M.}~\bibnamefont {Gieles}}, \bibinfo {author} {\bibfnamefont
  {E.}~\bibnamefont {Gosset}}, \bibinfo {author} {\bibfnamefont {R.~G.}\
  \bibnamefont {Izzard}}, \bibinfo {author} {\bibfnamefont {J.~B.~L.}\
  \bibnamefont {Bouquin}}, \ and\ \bibinfo {author} {\bibfnamefont {F.~R.~N.}\
  \bibnamefont {Schneider}},\ }\href {\doibase 10.1126/science.1223344}
  {\bibfield  {journal} {\bibinfo  {journal} {Science}\ }\textbf {\bibinfo
  {volume} {337}},\ \bibinfo {pages} {444} (\bibinfo {year} {2012})},\ \Eprint
  {http://arxiv.org/abs/1207.6397} {arXiv:1207.6397 [astro-ph.SR]} \BibitemShut
  {NoStop}%
\bibitem [{\citenamefont {{Iorio}}\ \emph {et~al.}(2023)\citenamefont
  {{Iorio}}, \citenamefont {{Mapelli}}, \citenamefont {{Costa}}, \citenamefont
  {{Spera}}, \citenamefont {{Escobar}}, \citenamefont {{Sgalletta}},
  \citenamefont {{Trani}}, \citenamefont {{Korb}}, \citenamefont
  {{Santoliquido}}, \citenamefont {{Dall'Amico}}, \citenamefont {{Gaspari}},\
  and\ \citenamefont {{Bressan}}}]{Iorio:2023sgz}%
  \BibitemOpen
  \bibfield  {author} {\bibinfo {author} {\bibfnamefont {G.}~\bibnamefont
  {{Iorio}}}, \bibinfo {author} {\bibfnamefont {M.}~\bibnamefont {{Mapelli}}},
  \bibinfo {author} {\bibfnamefont {G.}~\bibnamefont {{Costa}}}, \bibinfo
  {author} {\bibfnamefont {M.}~\bibnamefont {{Spera}}}, \bibinfo {author}
  {\bibfnamefont {G.~J.}\ \bibnamefont {{Escobar}}}, \bibinfo {author}
  {\bibfnamefont {C.}~\bibnamefont {{Sgalletta}}}, \bibinfo {author}
  {\bibfnamefont {A.~A.}\ \bibnamefont {{Trani}}}, \bibinfo {author}
  {\bibfnamefont {E.}~\bibnamefont {{Korb}}}, \bibinfo {author} {\bibfnamefont
  {F.}~\bibnamefont {{Santoliquido}}}, \bibinfo {author} {\bibfnamefont
  {M.}~\bibnamefont {{Dall'Amico}}}, \bibinfo {author} {\bibfnamefont
  {N.}~\bibnamefont {{Gaspari}}}, \ and\ \bibinfo {author} {\bibfnamefont
  {A.}~\bibnamefont {{Bressan}}},\ }\href {\doibase 10.1093/mnras/stad1630}
  {\bibfield  {journal} {\bibinfo  {journal} {\mnras}\ }\textbf {\bibinfo
  {volume} {524}},\ \bibinfo {pages} {426} (\bibinfo {year} {2023})},\ \Eprint
  {http://arxiv.org/abs/2211.11774} {arXiv:2211.11774 [astro-ph.HE]}
  \BibitemShut {NoStop}%
\bibitem [{\citenamefont {Kroupa}(2001)}]{Kroupa:2000iv}%
  \BibitemOpen
  \bibfield  {author} {\bibinfo {author} {\bibfnamefont {P.}~\bibnamefont
  {Kroupa}},\ }\href {\doibase 10.1046/j.1365-8711.2001.04022.x} {\bibfield
  {journal} {\bibinfo  {journal} {Mon. Not. Roy. Astron. Soc.}\ }\textbf
  {\bibinfo {volume} {322}},\ \bibinfo {pages} {231} (\bibinfo {year}
  {2001})},\ \Eprint {http://arxiv.org/abs/astro-ph/0009005}
  {arXiv:astro-ph/0009005} \BibitemShut {NoStop}%
\bibitem [{\citenamefont {Lagarde}\ \emph {et~al.}(2021)\citenamefont
  {Lagarde}, \citenamefont {R{\'e}yle}, \citenamefont {Chiappini},
  \citenamefont {Mor}, \citenamefont {Anders}, \citenamefont {Figueras},
  \citenamefont {Miglio}, \citenamefont {Romero-G{\'o}mez}, \citenamefont
  {Antoja}, \citenamefont {Cabral} \emph {et~al.}}]{lagarde2021deciphering}%
  \BibitemOpen
  \bibfield  {author} {\bibinfo {author} {\bibfnamefont {N.}~\bibnamefont
  {Lagarde}}, \bibinfo {author} {\bibfnamefont {C.}~\bibnamefont {R{\'e}yle}},
  \bibinfo {author} {\bibfnamefont {C.}~\bibnamefont {Chiappini}}, \bibinfo
  {author} {\bibfnamefont {R.}~\bibnamefont {Mor}}, \bibinfo {author}
  {\bibfnamefont {F.}~\bibnamefont {Anders}}, \bibinfo {author} {\bibfnamefont
  {F.}~\bibnamefont {Figueras}}, \bibinfo {author} {\bibfnamefont
  {A.}~\bibnamefont {Miglio}}, \bibinfo {author} {\bibfnamefont
  {M.}~\bibnamefont {Romero-G{\'o}mez}}, \bibinfo {author} {\bibfnamefont
  {T.}~\bibnamefont {Antoja}}, \bibinfo {author} {\bibfnamefont
  {N.}~\bibnamefont {Cabral}},  \emph {et~al.},\ }\href@noop {} {\bibfield
  {journal} {\bibinfo  {journal} {Astronomy \& Astrophysics}\ }\textbf
  {\bibinfo {volume} {654}},\ \bibinfo {pages} {A13} (\bibinfo {year}
  {2021})}\BibitemShut {NoStop}%
\bibitem [{\citenamefont {{Pinsonneault}}\ \emph {et~al.}(2018)\citenamefont
  {{Pinsonneault}}, \citenamefont {{Elsworth}}, \citenamefont {{Tayar}},
  \citenamefont {{Serenelli}}, \citenamefont {{Stello}}, \citenamefont
  {{Zinn}}, \citenamefont {{Mathur}}, \citenamefont {{Garc{\'\i}a}},
  \citenamefont {{Johnson}}, \citenamefont {{Hekker}}, \citenamefont {{Huber}},
  \citenamefont {{Kallinger}}, \citenamefont {{M{\'e}sz{\'a}ros}},
  \citenamefont {{Mosser}}, \citenamefont {{Stassun}}, \citenamefont
  {{Girardi}}, \citenamefont {{Rodrigues}}, \citenamefont {{Silva Aguirre}},
  \citenamefont {{An}}, \citenamefont {{Basu}}, \citenamefont {{Chaplin}},
  \citenamefont {{Corsaro}}, \citenamefont {{Cunha}}, \citenamefont
  {{Garc{\'\i}a-Hern{\'a}ndez}}, \citenamefont {{Holtzman}}, \citenamefont
  {{J{\"o}nsson}}, \citenamefont {{Shetrone}}, \citenamefont {{Smith}},
  \citenamefont {{Sobeck}}, \citenamefont {{Stringfellow}}, \citenamefont
  {{Zamora}}, \citenamefont {{Beers}}, \citenamefont
  {{Fern{\'a}ndez-Trincado}}, \citenamefont {{Frinchaboy}}, \citenamefont
  {{Hearty}},\ and\ \citenamefont {{Nitschelm}}}]{pinsonneault:2018}%
  \BibitemOpen
  \bibfield  {author} {\bibinfo {author} {\bibfnamefont {M.~H.}\ \bibnamefont
  {{Pinsonneault}}}, \bibinfo {author} {\bibfnamefont {Y.~P.}\ \bibnamefont
  {{Elsworth}}}, \bibinfo {author} {\bibfnamefont {J.}~\bibnamefont {{Tayar}}},
  \bibinfo {author} {\bibfnamefont {A.}~\bibnamefont {{Serenelli}}}, \bibinfo
  {author} {\bibfnamefont {D.}~\bibnamefont {{Stello}}}, \bibinfo {author}
  {\bibfnamefont {J.}~\bibnamefont {{Zinn}}}, \bibinfo {author} {\bibfnamefont
  {S.}~\bibnamefont {{Mathur}}}, \bibinfo {author} {\bibfnamefont {R.~A.}\
  \bibnamefont {{Garc{\'\i}a}}}, \bibinfo {author} {\bibfnamefont {J.~A.}\
  \bibnamefont {{Johnson}}}, \bibinfo {author} {\bibfnamefont {S.}~\bibnamefont
  {{Hekker}}}, \bibinfo {author} {\bibfnamefont {D.}~\bibnamefont {{Huber}}},
  \bibinfo {author} {\bibfnamefont {T.}~\bibnamefont {{Kallinger}}}, \bibinfo
  {author} {\bibfnamefont {S.}~\bibnamefont {{M{\'e}sz{\'a}ros}}}, \bibinfo
  {author} {\bibfnamefont {B.}~\bibnamefont {{Mosser}}}, \bibinfo {author}
  {\bibfnamefont {K.}~\bibnamefont {{Stassun}}}, \bibinfo {author}
  {\bibfnamefont {L.}~\bibnamefont {{Girardi}}}, \bibinfo {author}
  {\bibfnamefont {T.~S.}\ \bibnamefont {{Rodrigues}}}, \bibinfo {author}
  {\bibfnamefont {V.}~\bibnamefont {{Silva Aguirre}}}, \bibinfo {author}
  {\bibfnamefont {D.}~\bibnamefont {{An}}}, \bibinfo {author} {\bibfnamefont
  {S.}~\bibnamefont {{Basu}}}, \bibinfo {author} {\bibfnamefont {W.~J.}\
  \bibnamefont {{Chaplin}}}, \bibinfo {author} {\bibfnamefont {E.}~\bibnamefont
  {{Corsaro}}}, \bibinfo {author} {\bibfnamefont {K.}~\bibnamefont {{Cunha}}},
  \bibinfo {author} {\bibfnamefont {D.~A.}\ \bibnamefont
  {{Garc{\'\i}a-Hern{\'a}ndez}}}, \bibinfo {author} {\bibfnamefont
  {J.}~\bibnamefont {{Holtzman}}}, \bibinfo {author} {\bibfnamefont
  {H.}~\bibnamefont {{J{\"o}nsson}}}, \bibinfo {author} {\bibfnamefont
  {M.}~\bibnamefont {{Shetrone}}}, \bibinfo {author} {\bibfnamefont {V.~V.}\
  \bibnamefont {{Smith}}}, \bibinfo {author} {\bibfnamefont {J.~S.}\
  \bibnamefont {{Sobeck}}}, \bibinfo {author} {\bibfnamefont {G.~S.}\
  \bibnamefont {{Stringfellow}}}, \bibinfo {author} {\bibfnamefont
  {O.}~\bibnamefont {{Zamora}}}, \bibinfo {author} {\bibfnamefont {T.~C.}\
  \bibnamefont {{Beers}}}, \bibinfo {author} {\bibfnamefont {J.~G.}\
  \bibnamefont {{Fern{\'a}ndez-Trincado}}}, \bibinfo {author} {\bibfnamefont
  {P.~M.}\ \bibnamefont {{Frinchaboy}}}, \bibinfo {author} {\bibfnamefont
  {F.~R.}\ \bibnamefont {{Hearty}}}, \ and\ \bibinfo {author} {\bibfnamefont
  {C.}~\bibnamefont {{Nitschelm}}},\ }\href {\doibase 10.3847/1538-4365/aaebfd}
  {\bibfield  {journal} {\bibinfo  {journal} {\apjs}\ }\textbf {\bibinfo
  {volume} {239}},\ \bibinfo {eid} {32} (\bibinfo {year} {2018})},\ \Eprint
  {http://arxiv.org/abs/1804.09983} {arXiv:1804.09983 [astro-ph.SR]}
  \BibitemShut {NoStop}%
\bibitem [{\citenamefont {{Robin}}\ \emph {et~al.}(2003)\citenamefont
  {{Robin}}, \citenamefont {{Reyl{\'e}}}, \citenamefont {{Derri{\`e}re}},\ and\
  \citenamefont {{Picaud}}}]{robin:2003}%
  \BibitemOpen
  \bibfield  {author} {\bibinfo {author} {\bibfnamefont {A.~C.}\ \bibnamefont
  {{Robin}}}, \bibinfo {author} {\bibfnamefont {C.}~\bibnamefont
  {{Reyl{\'e}}}}, \bibinfo {author} {\bibfnamefont {S.}~\bibnamefont
  {{Derri{\`e}re}}}, \ and\ \bibinfo {author} {\bibfnamefont {S.}~\bibnamefont
  {{Picaud}}},\ }\href {\doibase 10.1051/0004-6361:20031117} {\bibfield
  {journal} {\bibinfo  {journal} {\aap}\ }\textbf {\bibinfo {volume} {409}},\
  \bibinfo {pages} {523} (\bibinfo {year} {2003})}\BibitemShut {NoStop}%
\bibitem [{\citenamefont {{Lagarde}}\ \emph {et~al.}(2017)\citenamefont
  {{Lagarde}}, \citenamefont {{Robin}}, \citenamefont {{Reyl{\'e}}},\ and\
  \citenamefont {{Nasello}}}]{lagarde:2017}%
  \BibitemOpen
  \bibfield  {author} {\bibinfo {author} {\bibfnamefont {N.}~\bibnamefont
  {{Lagarde}}}, \bibinfo {author} {\bibfnamefont {A.~C.}\ \bibnamefont
  {{Robin}}}, \bibinfo {author} {\bibfnamefont {C.}~\bibnamefont
  {{Reyl{\'e}}}}, \ and\ \bibinfo {author} {\bibfnamefont {G.}~\bibnamefont
  {{Nasello}}},\ }\href {\doibase 10.1051/0004-6361/201630253} {\bibfield
  {journal} {\bibinfo  {journal} {\aap}\ }\textbf {\bibinfo {volume} {601}},\
  \bibinfo {eid} {A27} (\bibinfo {year} {2017})},\ \Eprint
  {http://arxiv.org/abs/1702.01769} {arXiv:1702.01769 [astro-ph.SR]}
  \BibitemShut {NoStop}%
\bibitem [{\citenamefont {Fryer}\ \emph {et~al.}(2012)\citenamefont {Fryer},
  \citenamefont {Belczynski}, \citenamefont {Wiktorowicz}, \citenamefont
  {Dominik}, \citenamefont {Kalogera},\ and\ \citenamefont
  {Holz}}]{fryer2012compact}%
  \BibitemOpen
  \bibfield  {author} {\bibinfo {author} {\bibfnamefont {C.~L.}\ \bibnamefont
  {Fryer}}, \bibinfo {author} {\bibfnamefont {K.}~\bibnamefont {Belczynski}},
  \bibinfo {author} {\bibfnamefont {G.}~\bibnamefont {Wiktorowicz}}, \bibinfo
  {author} {\bibfnamefont {M.}~\bibnamefont {Dominik}}, \bibinfo {author}
  {\bibfnamefont {V.}~\bibnamefont {Kalogera}}, \ and\ \bibinfo {author}
  {\bibfnamefont {D.~E.}\ \bibnamefont {Holz}},\ }\href@noop {} {\bibfield
  {journal} {\bibinfo  {journal} {The Astrophysical Journal}\ }\textbf
  {\bibinfo {volume} {749}},\ \bibinfo {pages} {91} (\bibinfo {year}
  {2012})}\BibitemShut {NoStop}%
\bibitem [{\citenamefont {Belczynski}\ \emph {et~al.}(2020)\citenamefont
  {Belczynski} \emph {et~al.}}]{Belczynski:2017gds}%
  \BibitemOpen
  \bibfield  {author} {\bibinfo {author} {\bibfnamefont {K.}~\bibnamefont
  {Belczynski}} \emph {et~al.},\ }\href {\doibase 10.1051/0004-6361/201936528}
  {\bibfield  {journal} {\bibinfo  {journal} {Astron. Astrophys.}\ }\textbf
  {\bibinfo {volume} {636}},\ \bibinfo {pages} {A104} (\bibinfo {year}
  {2020})},\ \Eprint {http://arxiv.org/abs/1706.07053} {arXiv:1706.07053
  [astro-ph.HE]} \BibitemShut {NoStop}%
\bibitem [{\citenamefont {Bavera}\ \emph {et~al.}(2021)\citenamefont {Bavera},
  \citenamefont {Zevin},\ and\ \citenamefont {Fragos}}]{Bavera:2021evk}%
  \BibitemOpen
  \bibfield  {author} {\bibinfo {author} {\bibfnamefont {S.~S.}\ \bibnamefont
  {Bavera}}, \bibinfo {author} {\bibfnamefont {M.}~\bibnamefont {Zevin}}, \
  and\ \bibinfo {author} {\bibfnamefont {T.}~\bibnamefont {Fragos}},\
  }\href@noop {} {\  (\bibinfo {year} {2021})},\ \Eprint
  {http://arxiv.org/abs/2105.09077} {arXiv:2105.09077 [astro-ph.HE]}
  \BibitemShut {NoStop}%
\bibitem [{\citenamefont {Zevin}\ and\ \citenamefont
  {Bavera}(2022)}]{Zevin:2022wrw}%
  \BibitemOpen
  \bibfield  {author} {\bibinfo {author} {\bibfnamefont {M.}~\bibnamefont
  {Zevin}}\ and\ \bibinfo {author} {\bibfnamefont {S.~S.}\ \bibnamefont
  {Bavera}},\ }\href {\doibase 10.3847/1538-4357/ac6f5d} {\bibfield  {journal}
  {\bibinfo  {journal} {Astrophys. J.}\ }\textbf {\bibinfo {volume} {933}},\
  \bibinfo {pages} {86} (\bibinfo {year} {2022})},\ \Eprint
  {http://arxiv.org/abs/2203.02515} {arXiv:2203.02515 [astro-ph.HE]}
  \BibitemShut {NoStop}%
\bibitem [{\citenamefont {{Belczynski}}\ \emph {et~al.}(2013)\citenamefont
  {{Belczynski}}, \citenamefont {{Bulik}}, \citenamefont {{Mandel}},
  \citenamefont {{Sathyaprakash}}, \citenamefont {{Zdziarski}},\ and\
  \citenamefont {{Miko{\l}ajewska}}}]{Belczynski:2012}%
  \BibitemOpen
  \bibfield  {author} {\bibinfo {author} {\bibfnamefont {K.}~\bibnamefont
  {{Belczynski}}}, \bibinfo {author} {\bibfnamefont {T.}~\bibnamefont
  {{Bulik}}}, \bibinfo {author} {\bibfnamefont {I.}~\bibnamefont {{Mandel}}},
  \bibinfo {author} {\bibfnamefont {B.~S.}\ \bibnamefont {{Sathyaprakash}}},
  \bibinfo {author} {\bibfnamefont {A.~A.}\ \bibnamefont {{Zdziarski}}}, \ and\
  \bibinfo {author} {\bibfnamefont {J.}~\bibnamefont {{Miko{\l}ajewska}}},\
  }\href {\doibase 10.1088/0004-637X/764/1/96} {\bibfield  {journal} {\bibinfo
  {journal} {\apj}\ }\textbf {\bibinfo {volume} {764}},\ \bibinfo {eid} {96}
  (\bibinfo {year} {2013})},\ \Eprint {http://arxiv.org/abs/1209.2658}
  {arXiv:1209.2658 [astro-ph.HE]} \BibitemShut {NoStop}%
\bibitem [{\citenamefont {{Higgins}}\ \emph {et~al.}(2021)\citenamefont
  {{Higgins}}, \citenamefont {{Sander}}, \citenamefont {{Vink}},\ and\
  \citenamefont {{Hirschi}}}]{Higgins:2021}%
  \BibitemOpen
  \bibfield  {author} {\bibinfo {author} {\bibfnamefont {E.~R.}\ \bibnamefont
  {{Higgins}}}, \bibinfo {author} {\bibfnamefont {A.~A.~C.}\ \bibnamefont
  {{Sander}}}, \bibinfo {author} {\bibfnamefont {J.~S.}\ \bibnamefont
  {{Vink}}}, \ and\ \bibinfo {author} {\bibfnamefont {R.}~\bibnamefont
  {{Hirschi}}},\ }\href {\doibase 10.1093/mnras/stab1548} {\bibfield  {journal}
  {\bibinfo  {journal} {\mnras}\ }\textbf {\bibinfo {volume} {505}},\ \bibinfo
  {pages} {4874} (\bibinfo {year} {2021})},\ \Eprint
  {http://arxiv.org/abs/2105.12139} {arXiv:2105.12139 [astro-ph.SR]}
  \BibitemShut {NoStop}%
\bibitem [{\citenamefont {Kushnir}\ \emph {et~al.}(2017)\citenamefont
  {Kushnir}, \citenamefont {Zaldarriaga}, \citenamefont {Kollmeier},\ and\
  \citenamefont {Waldman}}]{Kushnir_2017}%
  \BibitemOpen
  \bibfield  {author} {\bibinfo {author} {\bibfnamefont {D.}~\bibnamefont
  {Kushnir}}, \bibinfo {author} {\bibfnamefont {M.}~\bibnamefont
  {Zaldarriaga}}, \bibinfo {author} {\bibfnamefont {J.~A.}\ \bibnamefont
  {Kollmeier}}, \ and\ \bibinfo {author} {\bibfnamefont {R.}~\bibnamefont
  {Waldman}},\ }\href {\doibase 10.1093/mnras/stx255} {\bibfield  {journal}
  {\bibinfo  {journal} {Monthly Notices of the Royal Astronomical Society}\
  }\textbf {\bibinfo {volume} {467}},\ \bibinfo {pages} {2146–2149} (\bibinfo
  {year} {2017})}\BibitemShut {NoStop}%
\bibitem [{\citenamefont {Qin}\ \emph {et~al.}(2018)\citenamefont {Qin},
  \citenamefont {Fragos}, \citenamefont {Meynet}, \citenamefont {Andrews},
  \citenamefont {S\o{}rensen},\ and\ \citenamefont {Song}}]{Qin:2018vaa}%
  \BibitemOpen
  \bibfield  {author} {\bibinfo {author} {\bibfnamefont {Y.}~\bibnamefont
  {Qin}}, \bibinfo {author} {\bibfnamefont {T.}~\bibnamefont {Fragos}},
  \bibinfo {author} {\bibfnamefont {G.}~\bibnamefont {Meynet}}, \bibinfo
  {author} {\bibfnamefont {J.}~\bibnamefont {Andrews}}, \bibinfo {author}
  {\bibfnamefont {M.}~\bibnamefont {S\o{}rensen}}, \ and\ \bibinfo {author}
  {\bibfnamefont {H.~F.}\ \bibnamefont {Song}},\ }\href {\doibase
  10.1051/0004-6361/201832839} {\bibfield  {journal} {\bibinfo  {journal}
  {Astron. Astrophys.}\ }\textbf {\bibinfo {volume} {616}},\ \bibinfo {pages}
  {A28} (\bibinfo {year} {2018})},\ \Eprint {http://arxiv.org/abs/1802.05738}
  {arXiv:1802.05738 [astro-ph.SR]} \BibitemShut {NoStop}%
\bibitem [{\citenamefont {Bavera}\ \emph {et~al.}(2020)\citenamefont {Bavera},
  \citenamefont {Fragos}, \citenamefont {Qin}, \citenamefont {Zapartas},
  \citenamefont {Neijssel}, \citenamefont {Mandel}, \citenamefont {Batta},
  \citenamefont {Gaebel}, \citenamefont {Kimball},\ and\ \citenamefont
  {Stevenson}}]{Bavera:2020inc}%
  \BibitemOpen
  \bibfield  {author} {\bibinfo {author} {\bibfnamefont {S.~S.}\ \bibnamefont
  {Bavera}}, \bibinfo {author} {\bibfnamefont {T.}~\bibnamefont {Fragos}},
  \bibinfo {author} {\bibfnamefont {Y.}~\bibnamefont {Qin}}, \bibinfo {author}
  {\bibfnamefont {E.}~\bibnamefont {Zapartas}}, \bibinfo {author}
  {\bibfnamefont {C.~J.}\ \bibnamefont {Neijssel}}, \bibinfo {author}
  {\bibfnamefont {I.}~\bibnamefont {Mandel}}, \bibinfo {author} {\bibfnamefont
  {A.}~\bibnamefont {Batta}}, \bibinfo {author} {\bibfnamefont {S.~M.}\
  \bibnamefont {Gaebel}}, \bibinfo {author} {\bibfnamefont {C.}~\bibnamefont
  {Kimball}}, \ and\ \bibinfo {author} {\bibfnamefont {S.}~\bibnamefont
  {Stevenson}},\ }\href {\doibase 10.1051/0004-6361/201936204} {\bibfield
  {journal} {\bibinfo  {journal} {Astron. Astrophys.}\ }\textbf {\bibinfo
  {volume} {635}},\ \bibinfo {pages} {A97} (\bibinfo {year} {2020})},\ \Eprint
  {http://arxiv.org/abs/1906.12257} {arXiv:1906.12257 [astro-ph.HE]}
  \BibitemShut {NoStop}%
\bibitem [{\citenamefont {Olejak}\ and\ \citenamefont
  {Belczynski}(2021)}]{Olejak:2021iux}%
  \BibitemOpen
  \bibfield  {author} {\bibinfo {author} {\bibfnamefont {A.}~\bibnamefont
  {Olejak}}\ and\ \bibinfo {author} {\bibfnamefont {K.}~\bibnamefont
  {Belczynski}},\ }\href {\doibase 10.3847/2041-8213/ac2f48} {\bibfield
  {journal} {\bibinfo  {journal} {Astrophys. J. Lett.}\ }\textbf {\bibinfo
  {volume} {921}},\ \bibinfo {pages} {L2} (\bibinfo {year} {2021})},\ \Eprint
  {http://arxiv.org/abs/2109.06872} {arXiv:2109.06872 [astro-ph.HE]}
  \BibitemShut {NoStop}%
\bibitem [{\citenamefont {Fuller}\ and\ \citenamefont
  {Lu}(2022)}]{Fuller:2022ysb}%
  \BibitemOpen
  \bibfield  {author} {\bibinfo {author} {\bibfnamefont {J.}~\bibnamefont
  {Fuller}}\ and\ \bibinfo {author} {\bibfnamefont {W.}~\bibnamefont {Lu}},\
  }\href {\doibase 10.1093/mnras/stac317} {\bibfield  {journal} {\bibinfo
  {journal} {Mon. Not. Roy. Astron. Soc.}\ }\textbf {\bibinfo {volume} {511}},\
  \bibinfo {pages} {3951} (\bibinfo {year} {2022})},\ \Eprint
  {http://arxiv.org/abs/2201.08407} {arXiv:2201.08407 [astro-ph.HE]}
  \BibitemShut {NoStop}%
\end{thebibliography}%

\appendix

\change{
\section{Checking the validity of treating the Newtonian potential as a perturbation}

In this appendix, we briefly check the validity of using the Newtonian potential and its higher-order multipole perturbations (e.g. \eqref{VstarMultipoleExpansion} and \eqref{Eq.DeltaGamma}) for our simulated dataset in this work.

First, it is important to note that due to the requirement of $\alpha\lesssim 0.4$, all participants are, by definition, non-relativistic. This means that at the cloud radius $r_n=n^2 r_1$, where $r_1=M/\alpha^2$ is the Bohr radius, the gravitational field of the Kerr BH is already weak enough to be treated by a Newtonian potential. Therefore, the validity of approximating the gravitational effect of the binary companion as a perturbation rests on the smallness of its Newtonian potential $\mathcal{V}_*$ compared to that of the BH.

We first consider the monopole component,
\begin{align}
    \mathcal{V}_*(r) = \frac{M_*}{R_*} \left[1+\mathcal{O}\left(\frac{r}{R_*}\right)\right]~,\quad R_*\sim p~,
\end{align}
where the binary separation $R_*$ varies with time in an orbital period and can be estimated by the semi-latus rectum. The Newtonian potential of the BH is
\begin{align}
    V(r)=\frac{M}{r} \left[1+\mathcal{O}\left(\frac{M}{r}\right)\right]~, \quad r\sim r_n=\frac{n^2 M}{\alpha^2}\gg M~.
\end{align}
The validity of perturbation theory requires
\begin{align}
    \frac{\mathcal{V}_*}{V}\sim q\times \frac{M}{p}\times\frac{n^2}{\alpha^2}\ll 1~,\label{perturbativityBound}
\end{align}
Thus if the mass ratio $q=M_*/M$ is overly large, the perturbative bound can be violated. We therefore scan our dataset for the participants and plot the distribution of the ratio $\mathcal{V}_*/V$ in Figure~\ref{fig:V_over}. Clearly all participants stay safely within the bound \eqref{perturbativityBound}.

Notice that both the monopole and the dipole can be removed by a coordinate transformation to the Fermi frame of the free-falling BH-cloud system \citep{Baumann:2018vus}. Thus the physical gravitational perturbation from the binary companion starts at quadrupolar order, with an additional suppression
\begin{align}
    \frac{V_*}{V}\sim \left(\frac{r_n}{p}\right)^2\times\frac{\mathcal{V}_*}{V}\ll \frac{\mathcal{V}_*}{V}\ll 1~,
\end{align}
further justifying the validity of perturbation theory.

The validity of treating $V_*$ as a perturbation also rests upon its smallness compared to the energy gap between the cloud modes, i.e. $V_* \ll \Delta E$, where $\Delta E$ is taken between the superradiant state and the absorptive state. For example, for the $\psi_{n22}$ and $\psi_{n00}$ mode we have 
\begin{equation}
\Delta E = E_{n22} - E_{n00} =\mu \left[\frac{72}{5} 
 \frac{\alpha^4}{n^4}+ O(\alpha^5)\right]~,
\end{equation}
To test if the gravitational perturbation
\begin{equation}
V_* \sim \mu \, q \, \frac{M_*}{R_*} \frac{r^2}{R_*^2} = \mu \, q \, \frac{M^3}{R_*^3} \frac{n^4}{\alpha^4}
\end{equation}
is much smaller than the free energy gap, we estimate the ratio by
\begin{equation}
\frac{V_*}{\Delta E} \sim \frac{5}{72} \, q \,\frac{M^3}{R_*^3} \frac{n^8}{\alpha^8}~,
\end{equation}
and plot the distribution of this ratio in Figure~\ref{fig:V_over}. The distribution shows that $V_*/\Delta E\ll 1$ always holds for the BH binaries in our simulation. This confirms that the perturbation potential remains much smaller than the energy gap, supporting the applicability of our previous approximation in this context.

\begin{figure}[htbp]
    \centering
    \includegraphics[width=0.4\textwidth]{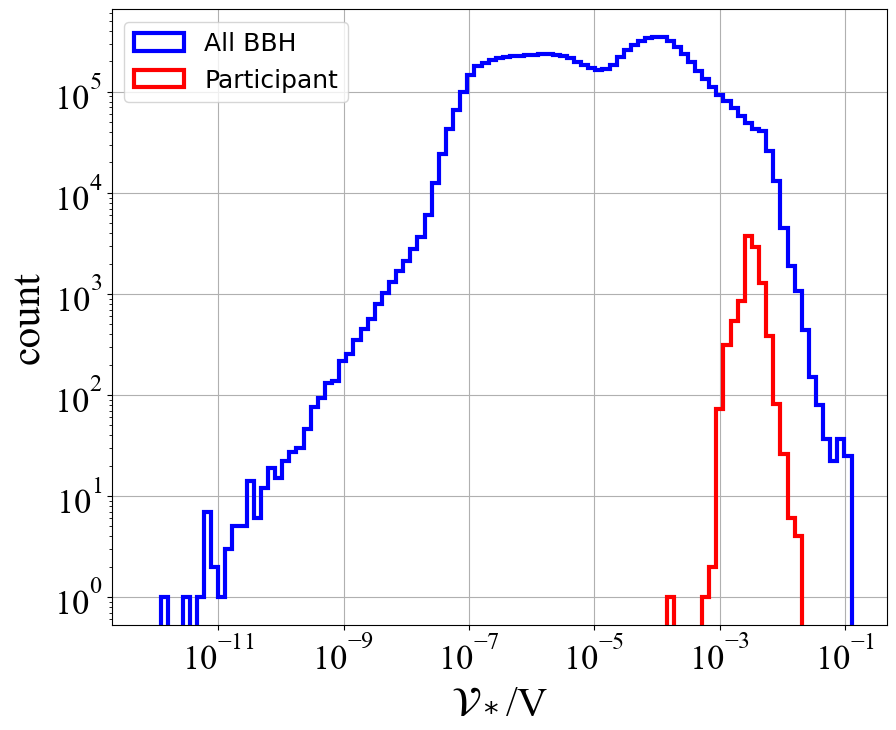}  \includegraphics[width=0.4\textwidth]{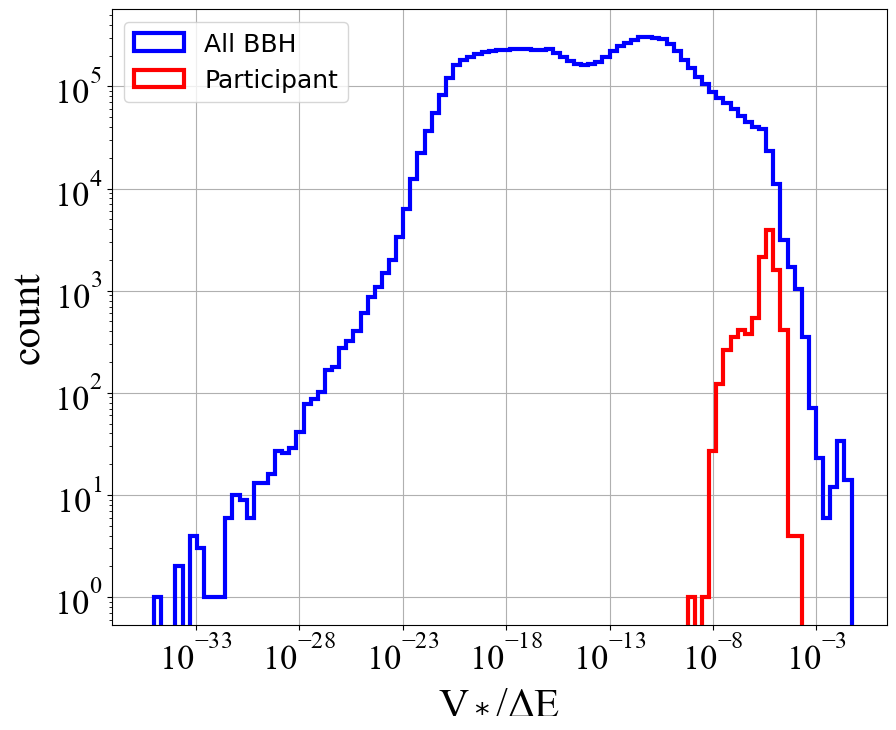}\\  
   \caption{
        Histogram of the ratios $\mathcal{V}_* / V$ \textit{(left)} and $V_* / \Delta E$ \textit{(right)} within our simulated dataset. The \textcolor{blue}{blue} curves are for all the BH binaries while the \textcolor{red}{red} curves are for participants only. Here we choose a reference boson mass $\mu=1.34\times10^{-12}$ eV. All systems are bounded by $\mathcal{V}_* / V<0.1$ and $V_*/\Delta E<0.06$, indicating that perturbativity is always under control. 
    }
    \label{fig:V_over}
\end{figure}

\noindent

}

\end{document}